\documentclass[aps,pre,reprint,amsmath,amssymb,superscriptaddress,showpacs,floatfix]{revtex4-1}
\usepackage{float}
\usepackage{graphicx}
\usepackage{dcolumn}
\usepackage{bm}
\usepackage[colorlinks, allcolors=blue]{hyperref}
\usepackage[usenames, dvipsnames]{color}
\begin{document}
\title{Machine learning approach to study quantum phases of a frustrated one dimensional spin-1/2 system}
\author{Sk Saniur Rahaman}
\email{saniur.rahaman@bose.res.in}
\author{Sumit Haldar}
\email{sumit.haldar@bose.res.in}
\author{Manoranjan Kumar}
\email{corresponding author: manoranjan.kumar@bose.res.in}
\affiliation{S. N. Bose National Centre for Basic Sciences, J D Block, Sector III, Salt Lake City, Kolkata 700106}
\date{\today}
\begin{abstract}
 Frustration driven quantum fluctuation leads to many exotic phases in the ground state and study of these quantum phase transitions is one of the most challenging areas of research in condensed matter physics. Here, a frustrated Heisenberg $J_1-J_2$ model of spin-1/2 chain with nearest exchange interaction $J_1$ and next nearest exchange interaction $J_2$ is studied using the principal component analysis (PCA) which is an unsupervised machine learning technique. In this method most probable spin configurations (MPSC) of ground-state (GS) and first excited state (FES) for different $J_2/J_1$ are used as the input in PCA to construct the co-variance matrix.
The `quantified principal component' of the largest eigenvalue of co-variance matrix $p_1(J_2/J_1)$ is calculated and it is shown that the nature and variation of $p_1(J_2/J_1)$ can accurately predict the phase transitions and degeneracies in the GS. The $p_1(J_2/J_1)$ calculated from the MPSC of GS can only exhibit the signature of degeneracies in the GS, whereas, $p_1(J_2/J_1)$ calculated from MPSC of FES captures the gapless spin liquid (GSL)-dimer phase transition as well as all the degeneracies of the model system. We show that jump in $p_1(J_2/J_1)$ of FES at $J_2/J_1 \approx 0.241$, indicates the GSL-dimer phase transition, whereas its kinks give the signature of the GS degeneracies. The scatter plot of first two principal components of FES shows distinct band formation for different phases.
\end{abstract}
\maketitle
\section{Introduction}
\label{sec:introduction}
Study of  quantum  fluctuations  and  phase transitions  is one of the most active and frontier area of condensed matter physics \cite{sachdev2011quantum,sachdev2008quantum,fazekas1999lecture,gdmahan,fetter2012quantum,si2010heavy,carr2010understanding,PhysRevB.102.024425,ckm1969,thouless2014quantum}, especially these studies are relevant and interesting in presence of many body interactions \cite{dutton2012108,zhitomirsky2010,chubukov1991,furukawa2012,parvej2017,saha2020modeling,maiti2019quantum}. These correlated model systems are important and relevant for modeling the real materials and have exotic phases in the ground state (GS), but solving even the simplest correlated model Hamiltonian is extremely difficult due to a large number of coupled degrees of freedom \cite{fazekas1999lecture,gdmahan,fetter2012quantum}. The degrees of freedom of a fermionic model system, like one-band Hubbard model increases as $4^N$, where $N$ is number of sites \cite{fazekas1999lecture,gdmahan,fetter2012quantum}. In large electronic repulsion energy limit, the charge degrees of freedom freeze and in half-filling limit, this model reduces to antiferromagnetic Heisenberg spin-1/2 model system where degrees of freedom increase as $2^N$ \cite{fazekas1999lecture,gdmahan,fetter2012quantum}. Unfortunately,  there are only a few models like one dimensional (1D) regular Heisenberg spin-1/2 \cite{bethe1931theorie,haldane1980general}  and Hubbard model \cite{takahashi1969magnetization,shiba1972magnetic} which can be solved exactly. In most of the cases, these models are solved by approximate analytical methods like mean-field \cite{affleck1988,marston1989}, bosonization theory \cite{giamarchi2003quantum}, renormalisation group theory \cite{shankar1994renormalization} etc or numerical methods like exact diagonalization (ED) for small systems \cite{soos1984vb,soos1983spin}, 
Density matrix renormalization group method (DMRG)
\cite{white-prl92,white-prb93,schollwock_dmrg,karen2006} or quantum Monte Carlo (QMC) \cite{sandvik2010,qmcex1} etc. Most of these techniques have their limitations like the DMRG can be applied only to low dimensional systems, the QMC have sign problems \cite{sandvik2010}  for frustrated spin systems or fermionic systems away from half filled limit. Therefore, it is desirable to find approach or method to determine the quantum phase boundary by supplying only minimal information like solving small system by using the ED method or giving some information about particle or spin configurations even  in presence of sign problems of the QMC method \cite{henelius2000}. The Machine learning (ML) is one of the most promising approach to find solution of complicated many body model Hamiltonian \cite{Carrasquilla,torlai2016learning,beach2018machine,broecker2017machine,shlens2014tutorial,rodriguez2019identifying,rem2019identifying,PhysRevB.96.195138}.

The ML is an indispensable tool to study various problems of different academic and social domain and has been heavily used in image recognition \cite{rosten2006machine,zoph2018learning,bishop2006pattern},  social networking \cite{liben2007link,islam2018depression,galan2016supervised}, advertising \cite{choimachine2020}, finance \cite{bose2001business}, 
designing medicine \cite{rajkomar2019machine} etc. In science, biological physics \cite{eskov2019heuristic,ding2018precision}, high energy physics \cite{albertsson2018machine,baldi2014searching} and other disciplines used this tool for various purposes.  In preponderance of information and real data, the ML is a excellent tool to explore and quantify patterns. There are two types of approaches in the ML; supervised and unsupervised learning, and in the supervised learning the algorithm is provided with any pre-assigned labels or scores for training the data 
\cite{Carrasquilla,shlens2014tutorial,torlai2016learning,guo2016deep}, whereas the unsupervised learning algorithm discover naturally occurring pattern in training data sets \cite{lecun2015deep,greplova2020unsupervised}. In this manuscript, the unsupervised learning will be our main focus to find the naturally occurring pattern.    

In a strongly correlated system, data are multivariate, highly correlated  and multi-dimensional, but to discover the pattern in the data set, the unsupervised learning first reduces the dimensionality without loosing the information of variance. Principal component analysis (PCA) is a method to reduce the dimensionality of a data set consisting of a large number of interrelated variables without loosing  variation in the given data set. The dimensionality reduction is achieved by transforming original data set to a new set of variables or the principal components (PCs) which are uncorrelated and ordered so that the first few axes of the subspace retain most of the variation present in the original variables \cite{jolliffe2002pca,jolliffe2016pca,woloshyn2019learning,shlens2014tutorial}. The PCA is one of the most efficient tool to  predict the thermal phase transitons in two dimesional (2D) spin systems like phase transition in classical  Ising model and XY model as a function of temperature $T$ \cite{PhysRevE.95.062122,PhysRevB.94.195105,PhysRevB.96.144432}.

The PCA method works well to determine the thermal phase transitions in model systems, but determination of quantum phase transition using the PCA is still  lacking in the literature. In this manuscript, we consider a 1D frustrated antiferromagnetic Heisenberg spin-1/2 $J_1-J_2$ model to explore the quantum phase transitions \cite{ckm1969,soos2016,okamoto1992,sirker2011,mkumar2015,mkumar2013quantum,parvej2017,saha2020modeling}. This model has nearest neighbour antiferromagnetic spin exchange interaction $J_1$  and  next nearest neighbour anti-ferromagnetic interaction $J_2$. This is one of the simplest frustrated model system irrespective of the nature of $J_2$ and one of the well studied model \cite{parvej2017,soos2016,okamoto1992,sirker2011,mkumar2015}. This model shows a gapless spin liquid (GSL) phase below  $J_2/J_1$=0.2411 \cite{chitra1995,mkumar2010} and a dimer phase thereafter. A spiral phase sets in at $J_2/J_1 \approx 0.55$. The quantum phase transition from the GSL phase to dimer phase is determined using level crossing between the first excited state (FES) singlet and lowest triplet states in finite systems or by calculating the singlet-triplet gap in thermodynamic limit \cite{mkumar2015}. This model is known as Majumdar-Ghosh (MG) model for $J_2/J_1=0.5$ and has a doubly degenerate GS \cite{ckm1969}. The GS of MG model can be written as  direct product of all dimers, and the lowest singlet triplet gap of  this model is large \cite{srwhite1996,chitra1995}. For $0.55 <J_2/J_1 <2.27 $, this model exhibits spiral phase and in this phase the GS is doubly degenerate only at specific values of $J_2/J_1$ for finite  size systems \cite{mkumar2013quantum}.

In this work, we explore the quantum phase transition and GS degeneracies in the $J_1-J_2$ model using PCA. We consider a few most probable spin configurations (MPSC) in the GS and the FES as input of the PCA for various values of $J_2/J_1$ and these MPSC are obtained using ED. We show that the MPSC corresponding to the FES for given value of $J_2/J_1$ predicts the GSL to dimer phase transition faithfully. It also predicts the degeneracies in the GS. It is amazing to see that only few MPSC with uniform weight can predict phase transitions and all the degeneracies. To best of our knowledge, the combined method of the PCA and the ED is first time used to predict the quantum phase transition and degeneracies.

This manuscript is divided into five sections. In section \ref{sec:model},
the model Hamiltonian and methods are discussed,
whereas in section \ref{sec:pca} the PCA and its implementation in this 
system are explained. The section \ref{sec:results} describes the results and have three subsections. In the last section \ref{sec:summary} our results  are summarised and compared with the results of the literature.
\section{Model Hamiltonian And Methods} 
\label{sec:model}
In this work, a frustrated Heisenberg antiferromagnetic spin-1/2 $J_1-J_2$ model is studied and in this model spins  are  interacting with it's nearest neighbour and next nearest neighbour spins with exchange  interactions $J_1$  and $J_2$ respectively. The Hamiltonian of the system can be written as
\begin{eqnarray}
\label{eqn:ham}
H=J_1\sum_{i}\vec{S}_i\cdot\vec{S}_{i+1}+J_2\sum_{i}\vec{S}_i\cdot\vec{S}_{i+2}
\end{eqnarray}
where $\vec{S_i}$ is the vector spin at site $i$, the first and second terms of the Hamiltonian represent the nearest and next nearest neighbour spin exchange interaction terms. Both the exchange interactions $J_1$ and $J_2$ are antiferromagnetic in nature. In large $J_2$ limit this interaction topology can be mapped to a Heisenberg model on a zigzag geometry or two spin chains coupled with each other through zigzag rung bonds \cite{chitra1995}.     

Conventional ED method is used to solve the Hamiltonian in Eq.[\ref{eqn:ham}]. The GS wavefunction $|\psi_0>$  and  the FES wavefunction  $|\psi_1>$ are calculated. The singlet-triplet (ST) gap  can be written as 
\begin{eqnarray}
\label{eqn:stgap}
E_{ST}=E_0(S^z=1)-E_0(S^z=0)
\end{eqnarray}
 where $E_0(S^z=0)$ and $E_0(S^z=1)$ are the lowest energy levels in $S^z=0$ and 1 sector i.e the lowest energy level in the singlet and triplet sector respectively. The lowest singlet excitation $E_{\sigma}$ can be written as  
 \begin{eqnarray}
\label{eqn:esigma}
E_{\sigma}=E_1(S=0)-E_0(S=0)
\end{eqnarray}
 where $E_1(S=0)$ and $E_0(S=0)$  are  the lowest and the FES energies in total spin $S=0$ sector respectively. For all the calculations, we impose the periodic boundary condition in the spin chain, and the ED is performed for three different system sizes $L=16$, $20$ and $24$. All the calculations are obtained for the wider regime  of $J_2/J_1$ from 0 to 1 in steps of 0.01.
\section{Principal Component Analysis}
\label{sec:pca}
In the PCA method, correlated data or features in a basis are rotated to 
a new basis which is linear combination of original basis while  
preserving all variation in the data \cite{jolliffe2002pca,jolliffe2016pca}. In the new rotated basis, most of the variations are confined to only a few dimensions and other dimensions are irrelevant. For determination of these important dimensions, we need the co-variance matrix of data sets of measurement. The co-variance matrix $C_T$ can be defined as 
\begin{equation}
\label{eqn:covar1}
C_T=XX^T
\end{equation}
 where, $X$ is sets of measurement with zero emperical mean  and 
it is a rectangular matrix. $X^T$ is transpose of $X$. We construct a data matrix $Y$ from which one can get the data-centred matrix $X$. Each row of $Y$  represents all measurements of a particular type, whereas column represents a set of measurements from one particular trial. In this case, $Y$ have $L$ dimensional features and $N_{J_2/J_1} =M \times m$ dimensional samples.
Where, $m$ is the number of configurations for each $J_2/J_1$ and $M$ is the number of equally spaced sets of $J_2/J_1$ for $0<J_2/J_1<1$. The data-centred matrix $X$ is formed by calculating the deviation of the components at each site of $Y$. For each site $i$ its mean $\mu_i$ is calculated by taking an average over all samples  and it is defined by ${\mu_i}=\frac{1}{N_{J_2/J_1}}\sum_{k=1}^{N_{J_2/J_1}} Y_{i,k}$. The deviation of the spin component for the $k^{th}$ sample at $i^{th}$ site can be written as $X_{i,k}=Y_{i,k}-{\mu_i}$. We write down the matrix elements of the co-variance matrix as 
\begin{equation}
\label{eqn:covar2}
C_T(i,j)=\sum_{k=1}^{N_{J_2/J_1}} X_{i,k}X^T_{k,j}. 
\end{equation}
$X^T_{k,j}$ is the transpose of $X_{j,k}$.

In thermal phase transition calculations, spin configuration at a particular 
temperature is a snap shot of a Monte-Carlo step \cite{PhysRevE.95.062122,PhysRevB.94.195105,PhysRevB.96.144432}. To study the quantum phase transition our first goal is to form the data matrix $Y$. In quantum phase transition, we deal with the wave function and it is defined as 
\begin{equation}
\label{eqn:psi}
|\psi\rangle=\sum_{k'=1}^{m} C_{k'}|\phi_{k'}\rangle
\end{equation}
where $|\phi_{k'}\rangle \equiv |\uparrow \downarrow \uparrow \downarrow  \downarrow ...  \rangle$ is an arbitrary spin configuration.  The configurations $ |\phi_{k'}\rangle$ are represented in $S^z$ basis and each configuration have information of $S^z$ component of each sites. $C_{k'}$ is amplitude and $C^2_{k'}$ is  the probability or weight factor of $ |\phi_{k'}\rangle$ in the wave function  as shown in Eq. \ref{eqn:psi}. We consider only  a few MPSC with respect to their relative weight, which is defined by
\begin{equation}
\label{eqn:relweig}
\gamma_{k'}=\frac{C^2_{k'}}{\operatorname{Max[C^2_1,...,C^2_{N_H}]}}
\end{equation}
$N_H$ is the dimension of the Hilbert's space. The largest value of $\gamma_{k'}$ is  unity and subsequent relative weight decreases as probability decreases. In our present study, we consider first $m$ fixed number  of most probable spin configuration $|\phi_{k'}\rangle$ for each value of $J_2/J_1$ and now $L \times N_{J_2/J_1}$ 
dimensional matrix X is constructed. For a system size $L$, $C^T$ have dimension of $L \times L$ and there are $L$ eigenvalues [$\mathbf{\lambda_1},...,\mathbf{\lambda_L}$] and its corresponding eigen vectors or weight vectors [$\mathbf{w_1},...,\mathbf{w_L}$] after diagonalisation. In general one can write,
\begin{equation}
\mathbf{X}\mathbf{X^T} \mathbf{w_n} = \lambda_n \mathbf{w_n}
\end{equation}
where, each eigenvector $w_n$ is a column vector with $L$ rows. The variance of the data set for various $J_2/J_1$ is maximum corresponds to the largest 
eigenvalue $\lambda_1$ and it is second maximum for the second largest eigenvalue $\lambda_2$ and so on. Using such dimensionality 
reduction procedure, we find only few large eigenvalues and corresponding eigenvectors are important to accommodate most of the variation in the original data sets. 

By projecting the original data along $n^{th}$ eigenvector $w_n$, we get $N_{J_2/J_1}$ number of principal components. The $n^{th}$ principal component $p_n(J_2/J_1, k)$ corresponds to the $k^{th}$ sample
can be obtained by-
\begin{equation}
\label{eqn:pcn}
p_n(J_2/J_1,k)=w_n^T Y(k)
\end{equation}
here, $Y(k)$ is the $k^{th}$ sample with $L$ entries of the data matrix $Y$. $w^T_n$ is the transpose of the eigenvector $w_n$. The $n^{th}$ `quantified principal component'  is calculated by summing the absolute value of the principal components over the samples
($m$) for a particular value of $J_2/J_1$ \cite{PhysRevE.95.062122}.
\begin{equation}
\label{eqn:qpc}
p_n(J_2/J_1)=\sum_{k=1}^{m} |p_n(J_2/J_1,k)|
\end{equation}

In this manuscript, we focus on the PCA of GS and the FES of the system for different $J_2/J_1$ and calculate the eigenvalues and the corresponding eigenvectors of co-variance matrix $C_T$. We calculate principal components corresponding to the largest eigenvalues. We note that in our study, first eigenvalue is the largest as compared to the others and so the `quantified principal component' $p_1(J_2/J_1)$  is able to determine the phase transition, and our calculations also suggest that the principal components of the FES is more important in determining the quantum phase transition.

\section{Results}
\label{sec:results}
In this section, we first discuss quantum phase transition and phases boundary of results of $J_1-J_2$ model in the literature, and thereafter, we apply a unsupervised mechine learning method to evaluate quantum phase boundary and degeneracies in this model. The quantum  phase transiton of $J_1-J_2$ model for a spin-1/2 has been extensively studied by using different analytical and numerical techniques \cite{soos2016,sirker2011,mkumar2015}. In small $J_2/J_1$ limit, the GS exhibits the GSL phase which is characterized by algebraic decay of spin correlation function and gapless spectrum \cite{mkumar2010,srwhite1996,chitra1995}. This phase extends upto $J_2/J_1$=0.2411, and thereafter, a gapped  phase sets in the GS  and spin correlations are short range or exponentially decaying \cite{mkumar2010,srwhite1996,chitra1995}. In this phase, the system shows dimerized GS and finite $E_{ST}$. $E_{ST}$ and nature of spin correlation function can be used to determine the phase boundary in the thermodynamic limit 
\cite{okamoto1992, sirker2011, mkumar2015}. In a finite system, crossover point of the FES of singlet and the lowest state of triplet is phase boundary of the GSL and the dimer phase \cite{mkumar2015}. 

Now let us apply an unsupervised machine learning technique PCA to determine the phase bourndary and signature of degeneracies in the GS. We pointed out in section \ref{sec:pca} that the eigenvalues and the eigenvectors of $C^T$ are important to decide the rotation of  basis along which maximum of the variation is preserved. Only a few  largest eigenvalues and their eigenvectors are important to determine relevant principal components. In the rotated basis, $n^{th}$ principal component  corresponds to the  $k^{th}$ sample ${p_n(J_2/J_1, k)}$ can be obtained from Eq. \ref{eqn:pcn} and the `quantified principal component' $p_n(J_2/J_1)$  from Eq. \ref{eqn:qpc}. We calculate the  `quantified principal component' $p_1(J_2/J_1)$ of the GS and the FES in $S^z=0$ sector to characterize the phase boundary of this model. The data centred-matrix $X$ is calculated using MPSC approach as described in section \ref{sec:pca}. 
 
The eigenvalues of $C^T$ are analysed and the effect of MPSC $m$ and 
the effect of finite size $L$ on principal components are studied in detail. The phase boundary and the degeneracies are studied with the help of `quantified principal component' $p_1(J_2/J_1)$ corresponding to the largest eigenvalue $\lambda_1$. In all our calculations, we have taken $M=100$ sets of $J_2/J_1$ in the range $0<J_2/J_1<1$ with step $0.01$.

\subsection{Effect of $L$ and $m$ on $\lambda_n$} 
\begin{figure}[t]
\includegraphics[width=0.99\linewidth]{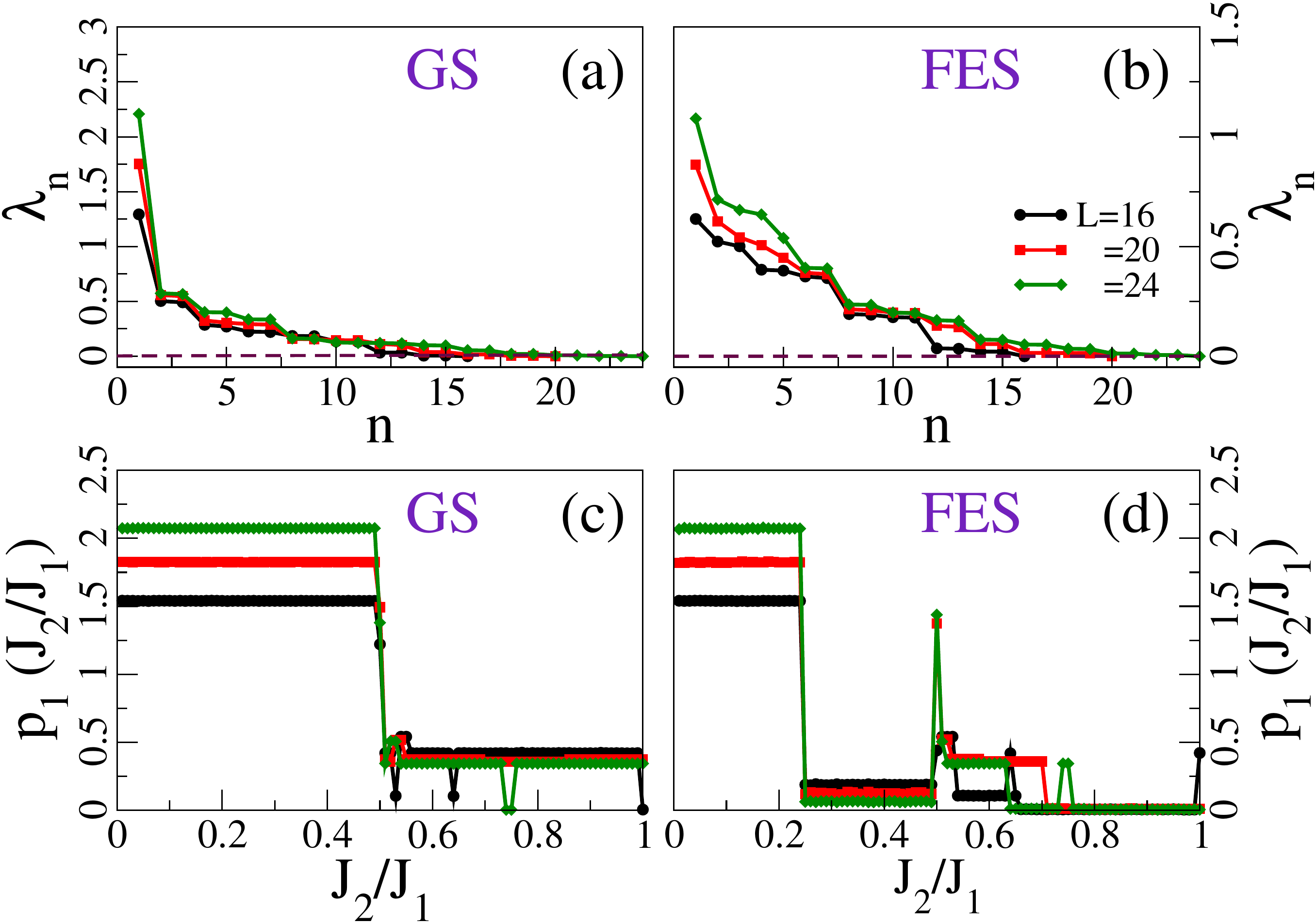}
\caption{(color online) Finite size effect on `quantified principal component' $p_1(J_2/J_1)$ is shown for $m=25$. Eigenvalues of the co-variance matrix are shown for (a) the GS and (b) the FES. `quantified principal component' $p_1(J_2/J_1)$ are shown for (c) the GS and (d) the FES. Black, red and green colors represent different system sizes $L=16, 20$ and $24$ respectively.}
\label{fig:diff_size}
\end{figure}

The eigenvalues $\lambda_n$ of the co-variance matrix for the GS and the FES are shown in  Fig. \ref{fig:diff_size} for three different system sizes $L=16$, $20$ and $24$ for $m=25$. For fixed $m=25$, $\lambda_n$ of the GS and FES decrease rapidly as shown in Fig. \ref{fig:diff_size}(a) and \ref{fig:diff_size}(b) respectively. In both the cases, the first eigenvalue is dominating over others and increases with system size for both GS and FES. Therefore, we choose the first eigenvalue and its corresponding `quantified principal component' $p_1(J_2/J_1)$. In Fig. \ref{fig:diff_size}(c) and \ref{fig:diff_size}(d), we plot $p_1(J_2/J_1)$ for the GS and the FES respectively as a function of $J_2/J_1$ for three different system sizes $L$. 
We observe $p_1(J_2/J_1)$ is constant and suddenly drops at $J_2/J_1 = 0.5$ (MG point) for GS. However, for the FES, we notice a drop at $J_2/J_1 = 0.241$ and thereafter it is constant upto $J_2/J_1 = 0.5$ for all values of $L$. We also notice few more in between $0.5\leq J_2/J_1 \leq 1.0$. In GS, the jumps are observed for  $0.55<J_2/J_1<1.0$ for the system sizes $L=16$, $24$ respectively. However, in case of the FES, we notice two jumps for all the system sizes. The values of $p_1(J_2/J_1)$ below $J_2/J_1<0.5$ increases with $L$, whereas it decreases with $L$ for $J_2/J_1>0.5$. 
\begin{figure}[t]
\includegraphics[width=0.99\linewidth]{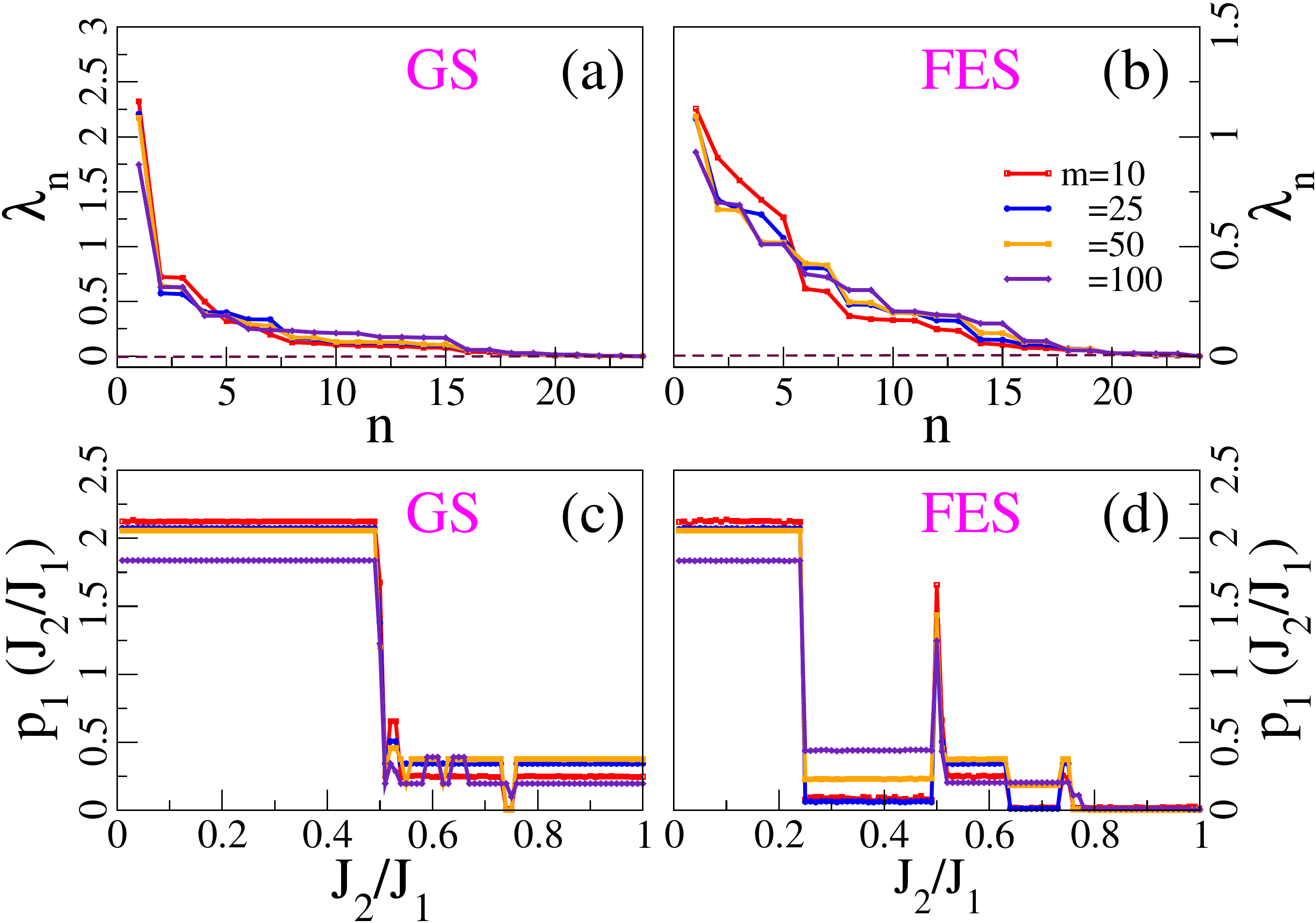}
\caption{(color online) PCA dependence on MPSC ($m$) taken for the data matrix for $L=24$ is shown. Eigenvalues of the co-variance matrix are shown for (a) the GS and (b) the FES and `quantified principal component' $p_1(J_2/J_1)$ are shown for (c) the GS and (d) the FES. Red ,blue, orange, indigo colors correspond to $m=10$, $25$, $50$ and $100$ respectively.}
\label{fig:diff_diag}
\end{figure}
\begin{figure}
\includegraphics[width=0.99\linewidth]{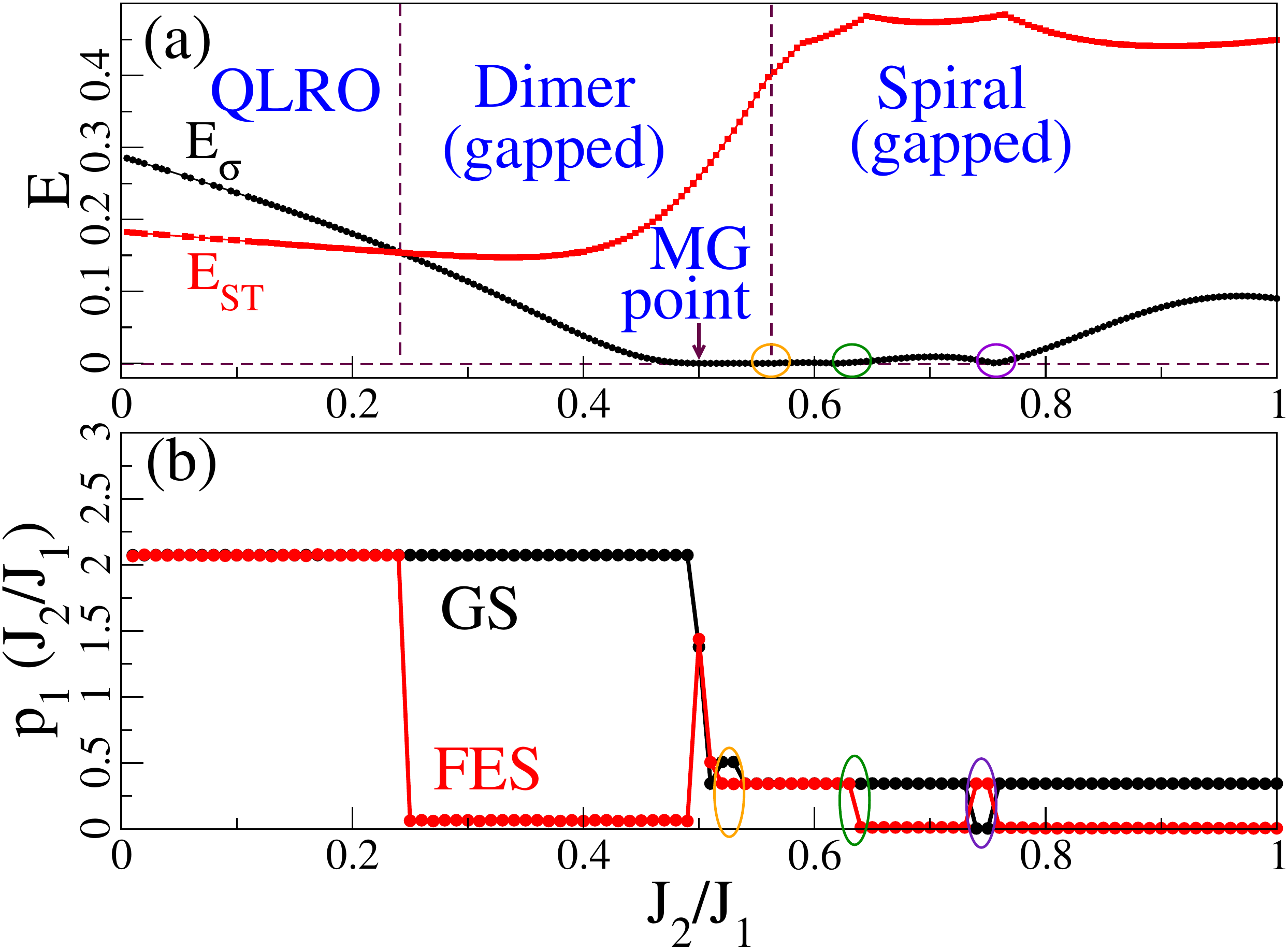}
\caption{(color online) For $L=24$ (a)  Energy gaps vs $J_2/J_1$ is plotted. Black curve is first excited state energy gap ($E_\sigma$) defined in Eq. \ref{eqn:stgap} and red curve is the lowest triplet gap ($E_{ST}$) defined in Eq. \ref{eqn:stgap}. (b) `quantified principal component' $p_1(J_2/J_1)$ is shown as a function of $J_2/J_1$. Black and red curves represent $p_1(J_2/J_1)$ calculated from the GS and the FES respectively.}
\label{em_pr}
\end{figure}

Now we analyse the dependence of $\lambda_n$ and $p_1(J_2/J_1)$ on number of sample $m$  for both the GS and the FES. Four different values $m=10$, $25$, $50$ and $100$ are considered as shown in  Fig. \ref{fig:diff_diag} for  $L=24$. $\lambda_n$ and their variation in both the GS and FES decreases with $m$ as shown in Fig. \ref{fig:diff_diag}(a) and \ref{fig:diff_diag}(b) respectively. In fact, MPSC are governed by the Hamiltonian and increasing the $m$ increases the chance of inclusion of even less probable configurations which hardly contributes to the phase transition. Only first `quantified principal component' $p_1(J_2/J_1)$ is important in both the GS and the FES. In Fig. \ref{fig:diff_diag}(c) and \ref{fig:diff_diag}(d), $p_1(J_2/J_1)$ as a function of $J_2/J_1$ is plotted for the GS and FES respectively for four values of $m$. The expected jumps and fluctuations in $p_1(J_2/J_1)$ get suppressed with increasing $m$, but their nature is quite similar for  all values of $m$.  We present all the results for $m=25$ hereafter. 

\begin{figure*}
\includegraphics[width=0.99\linewidth]{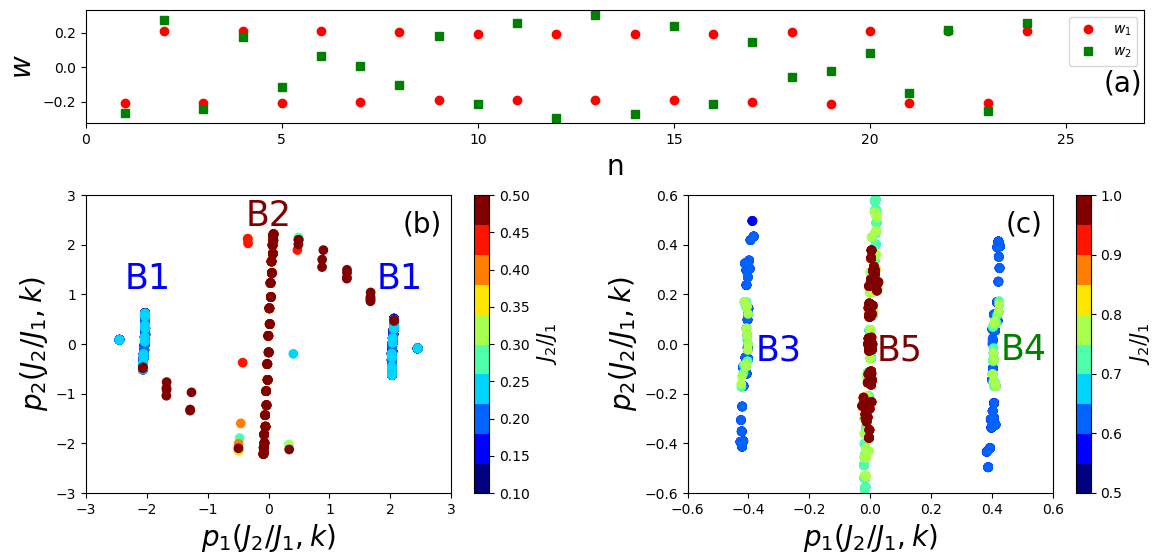}
\caption{(color online) For $L=24$, $m=25$: (a) eigen vectors $\mathbf{w_1}$ and $\mathbf{w_2}$ corresponding to largest and second largest eigenvalues of $C^T$ are plotted as a function of site $n$. (b) Scatter plot of $p_1(J_2/J_1)$ vs $p_2(J_2/J_1)$ for the limit  $0.1<J_2/J_1<0.5$. The color bands B1 (sky blue) and B2 (maroon) are separated at $J_2/J_1 \approx 0.241$. (c) Scatter plot of $p_1(J_2/J_1)$ vs $p_2(J_2/J_1)$ for the limit  $0.5<J_2/J_1<1.0$. The color bands B3 (blue), B4 (green) are separated around $J_2/J_1=0.63$ and B4, B5 (maroon) are separated at $J_2/J_1=0.745$.}
\label{fig:scatter}
\end{figure*}

\subsection{Principal components and quantum phase transition}
In this work, we are dealing with the finite system size $L=24$, therefore let us try to compare the $p_1(J_2/J_1)$ curve with $E_{ST}-E_{\sigma}$ curve as a function of $J_2/J_1$. In Fig. \ref{em_pr}(a), $E_{ST}-E_{\sigma}$ curve is plotted as a function of $J_2/J_1$, there is a crossing of $E_{ST}$ and  $E_{\sigma}$ curve at $J_2/J_1 \approx 0.241$ and this crossing indicates the GSL-Dimer phase boundary for $L=24$. At $J_2/J_1=0.5$ or MG point and $0.55$, $0.62$, $0.755$, the $E_{\sigma}$ is zero i.e  ground state is doubly degenerate as shown in Fig. \ref{em_pr}(a). The degeneracies are marked with circles. In Fig. \ref{em_pr}(b), $p_1(J_2/J_1)$ vs $J_2/J_1$ for the GS, exhibits a large jump at $J_2/J_1=0.5$ and kinks at $0.52$, $0.745$,  however, near to the MG point, fluctuations in $p_1(J_2/J_1)$ curve indicate degeneracies or nearly degenerate states. The GS $p_1(J_2/J_1)$ does not show any kink near $0.62$. The $p_1(J_2/J_1)$ of the FES vs $J_2/J_1$ exhibits a large jump at $J_2/J_1 \approx 0.241$, jump at MG point and kinks at $J_2/J_1=0.52$, $0.63$, $0.745$. It is well known that degeneracies at $0.55$, $0.62$, $0.755$ are due to incommensurate phases i.e, two singlet states become degenerate at these points \cite{mkumar2013quantum}. Therefore, we can conclude that by studying  the `quantified principal component' $p_1(J_2/J_1)$ of the GS and FES, one can extract the degeneracies in the GS, however, analysis of the $p_1(J_2/J_1)$ of the FES can reliably give phase boundary of the GSL-dimer phase transition and these results are consistent with results in the literature \cite{mkumar2013quantum,mkumar2015,chitra1995,srwhite1996}.       

\subsection{Principal components and eigenvectors of $C_T$}
The co-variance matrix $C_T$ is calculated from the FES wave function 
for $0 <J_2/J_1< 1$ and first two eigenvalues $\lambda_1$ and $\lambda_2$ of 
$C_T$ have significantly larger variances as compared to the other eigenvalues. The eigenvectors $w_1$ and $w_2$ corresponding to $\lambda_1$ and $\lambda_2$ are plotted in Fig. \ref{fig:scatter}(a), and we notice that $w_1$ is a periodic function with pitch angle $\pi$ where $w_2$ is a periodic function with pitch angle $\pi/6$. Now, $p_1(J_2/J_1,k)$ and $p_2(J_2/J_1,k)$  can be constructed by rotating the original $Y$ by these two weightvectors $w_1$ and 
$w_2$ by using expression Eq. \ref{eqn:pcn}. The  scatter color plot of $p_1(J_2/J_1,k)$ and $p_2(J_2/J_1,k)$ are shown in Fig. \ref{fig:scatter} (b) and \ref{fig:scatter}(c), and the color intensity bar indicate the strength of $J_2/J_1$. The $p_1(J_2/J_1,k)$ and $p_2(J_2/J_1,k)$ are plotted for $0.1<J_2/J_1 <0.5$ in Fig. \ref{fig:scatter}(b), which form two kinds of bands B1 and B2. The principal components form band B1 in the range  $0.1 \leq J_2/J_1 \leq 0.241$ and this band is located at $p_1(J_2/J_1,k) \approx \pm 2.0$. For $0.241 \leq J_2/J_1 \leq 0.5$, principal components form B2 band around $p_1(J_2/J_1,k)=0$ and a clear separation of data can be seen around $J_2/J_1=0.241$.

In Fig. \ref{fig:scatter}(c), the scatter plot is shown for $0.5< J_2/J_1 <1$ and the principal components form three color bands. These bands B3, B4, B5 correspond to the limits $0.5 \leq J_2/J_1 \leq 0.63$, $0.63 \leq J_2/J_1 \leq 0.745$ and $0.745 \leq J_2/J_1 \leq 1.0$ respectively. These bands correspond to different pitch angle in the system in spiral phase. The band B3 are located at $p_1(J_2/J_1,k) \approx \pm 0.4$ and represented with blue color, B4 is  symmetrically located at $|p_1(J_2/J_1,k)| \approx 0$ and $|p_1(J_2/J_1,k)| \approx 0.4$. The major fraction of the components of B4 reside at $|p_1(J_2/J_1,k)| \approx 0$ rather $|p_1(J_2/J_1,k)| \approx 0.4$ and so the average of it gives finite but near equals to zero value of $|p_1(J_2/J_1,k)|$ as shown in Fig. \ref{em_pr}(b). The band B5 is located along $|p_1(J_2/J_1,k)| \approx 0$ and it gives the average $|p_1(J_2/J_1,k)|=0$ as shown in Fig. \ref{em_pr}(b).         

\section{Summary}
\label{sec:summary}
In this manuscript, a one dimensional antiferromagnetic Heisenberg spin-1/2 $J_1-J_2$ model is considered to study the quantum phase transition and the GS degeneracies using the machine learning method. We have combined the  ED and 
the PCA to study the  quantum phase transition boundaries and degeneracies of the model system in Eq. \ref{eqn:ham}, this method can predict the phase boundaries by using finite size calculations. The $J_1-J_2$ model has the GSL-dimer phase boundary at $J_2/J_1=0.2411$ calculated by DMRG method \cite{chitra1995,srwhite1996} and finite systems shows GS degeneracies in the spiral phase \cite{mkumar2013quantum}. In this manuscript, it is shown that the  quantum phase boundaries and degeneracies in finite system are predicted by using the PCA  considering only few number of MPSC in the GS and the FES of the model system. This indeed a promising tool to capture the quantum fluctuation and correlation only taking few most probable configurations out of larger dimensional Hilbert's space. 

We show that the nature of the `quantified principal component' $p_1(J_2/J_1)$ is not strongly sensitive to variation  of $m$  and the sensitivity is shown for $m=10$, $25$, $50$ and $100$ in Fig. \ref{fig:diff_diag}. The nature of $p_1(J_2/J_1)$ of the FES also does not change with system size and its variation at the boundary gets enhanced with increasing $L$. The jumps and kinks in $p_1(J_2/J_1)$ can give an accurate determination of phase boundary and degeneracies in the GS. The PCA of the GS can predict some of the degeneracies in the GS. However, $p_1(J_2/J_1)$ of the FES exhibits jumps at $J_2/J_1=0.241$ and this jump represents the GSL-dimer phase boundary and this phase boundary is consistent with literature  and the boundary has been determined by $E_{ST}$ calculation in thermodynamic limit \cite{soos2016} and using level crossing method for finite system size \cite{mkumar2015}. The kinks at  $J_2/J_1=0.52, 0.63, 0.745$ for $L=24$ represent the degeneracies in the GS and these are consistent with literature \cite{mkumar2013quantum}. In conclusion, a combined method of ED and unsupervised machine learning method PCA gives reliable phase boundaries  and this method can be a powerful tool to explore the quantum phase transition in strongly correlated systems. 
 
\section{Acknowledgements}
M.K. thanks SERB for financial support through GrantSanction No. CRG/2020/000754. S.S.R. thanks University 
Grants Commmision (UGC) for fellowship.
\bibliography{ref_ladder}      

\begin{thebibliography}{72}%
\makeatletter
\providecommand \@ifxundefined [1]{%
 \@ifx{#1\undefined}
}%
\providecommand \@ifnum [1]{%
 \ifnum #1\expandafter \@firstoftwo
 \else \expandafter \@secondoftwo
 \fi
}%
\providecommand \@ifx [1]{%
 \ifx #1\expandafter \@firstoftwo
 \else \expandafter \@secondoftwo
 \fi
}%
\providecommand \natexlab [1]{#1}%
\providecommand \enquote  [1]{``#1''}%
\providecommand \bibnamefont  [1]{#1}%
\providecommand \bibfnamefont [1]{#1}%
\providecommand \citenamefont [1]{#1}%
\providecommand \href@noop [0]{\@secondoftwo}%
\providecommand \href [0]{\begingroup \@sanitize@url \@href}%
\providecommand \@href[1]{\@@startlink{#1}\@@href}%
\providecommand \@@href[1]{\endgroup#1\@@endlink}%
\providecommand \@sanitize@url [0]{\catcode `\\12\catcode `\$12\catcode
  `\&12\catcode `\#12\catcode `\^12\catcode `\_12\catcode `\%12\relax}%
\providecommand \@@startlink[1]{}%
\providecommand \@@endlink[0]{}%
\providecommand \url  [0]{\begingroup\@sanitize@url \@url }%
\providecommand \@url [1]{\endgroup\@href {#1}{\urlprefix }}%
\providecommand \urlprefix  [0]{URL }%
\providecommand \Eprint [0]{\href }%
\providecommand \doibase [0]{http://dx.doi.org/}%
\providecommand \selectlanguage [0]{\@gobble}%
\providecommand \bibinfo  [0]{\@secondoftwo}%
\providecommand \bibfield  [0]{\@secondoftwo}%
\providecommand \translation [1]{[#1]}%
\providecommand \BibitemOpen [0]{}%
\providecommand \bibitemStop [0]{}%
\providecommand \bibitemNoStop [0]{.\EOS\space}%
\providecommand \EOS [0]{\spacefactor3000\relax}%
\providecommand \BibitemShut  [1]{\csname bibitem#1\endcsname}%
\let\auto@bib@innerbib\@empty
\bibitem [{\citenamefont {Sachdev}(2011)}]{sachdev2011quantum}%
  \BibitemOpen
  \bibfield  {author} {\bibinfo {author} {\bibfnamefont {S.}~\bibnamefont
  {Sachdev}},\ }\href@noop {} {\emph {\bibinfo {title} {Quantum phase
  transitions}}}\ (\bibinfo  {publisher} {Cambridge university press},\
  \bibinfo {year} {2011})\BibitemShut {NoStop}%
\bibitem [{\citenamefont {Sachdev}(2008)}]{sachdev2008quantum}%
  \BibitemOpen
  \bibfield  {author} {\bibinfo {author} {\bibfnamefont {S.}~\bibnamefont
  {Sachdev}},\ }\href {https://doi.org/10.1038/nphys894} {\bibfield  {journal}
  {\bibinfo  {journal} {Nature Physics}\ }\textbf {\bibinfo {volume} {4}},\
  \bibinfo {pages} {173} (\bibinfo {year} {2008})}\BibitemShut {NoStop}%
\bibitem [{\citenamefont {Fazekas}(1999)}]{fazekas1999lecture}%
  \BibitemOpen
  \bibfield  {author} {\bibinfo {author} {\bibfnamefont {P.}~\bibnamefont
  {Fazekas}},\ }\href@noop {} {\emph {\bibinfo {title} {Lecture notes on
  electron correlation and magnetism}}},\ Vol.~\bibinfo {volume} {5}\ (\bibinfo
   {publisher} {World scientific},\ \bibinfo {year} {1999})\BibitemShut
  {NoStop}%
\bibitem [{\citenamefont {Mahan}(2000)}]{gdmahan}%
  \BibitemOpen
  \bibfield  {author} {\bibinfo {author} {\bibfnamefont {G.~D.}\ \bibnamefont
  {Mahan}},\ }\href {\doibase 10.1007/978-1-4757-5714-9} {\emph {\bibinfo
  {title} {Many-Particle Physics}}},\ \bibinfo {edition} {3rd}\ ed.\ (\bibinfo
  {publisher} {Springer US},\ \bibinfo {year} {2000})\BibitemShut {NoStop}%
\bibitem [{\citenamefont {Fetter}\ and\ \citenamefont
  {Walecka}(2012)}]{fetter2012quantum}%
  \BibitemOpen
  \bibfield  {author} {\bibinfo {author} {\bibfnamefont {A.~L.}\ \bibnamefont
  {Fetter}}\ and\ \bibinfo {author} {\bibfnamefont {J.~D.}\ \bibnamefont
  {Walecka}},\ }\href@noop {} {\emph {\bibinfo {title} {Quantum theory of
  many-particle systems}}}\ (\bibinfo  {publisher} {Courier Corporation},\
  \bibinfo {year} {2012})\BibitemShut {NoStop}%
\bibitem [{\citenamefont {Si}\ and\ \citenamefont
  {Steglich}(2010)}]{si2010heavy}%
  \BibitemOpen
  \bibfield  {author} {\bibinfo {author} {\bibfnamefont {Q.}~\bibnamefont
  {Si}}\ and\ \bibinfo {author} {\bibfnamefont {F.}~\bibnamefont {Steglich}},\
  }\href {\doibase 10.1126/science.1191195} {\bibfield  {journal} {\bibinfo
  {journal} {Science}\ }\textbf {\bibinfo {volume} {329}},\ \bibinfo {pages}
  {1161} (\bibinfo {year} {2010})}\BibitemShut {NoStop}%
\bibitem [{\citenamefont {Carr}(2010)}]{carr2010understanding}%
  \BibitemOpen
  \bibfield  {author} {\bibinfo {author} {\bibfnamefont {L.}~\bibnamefont
  {Carr}},\ }\href@noop {} {\emph {\bibinfo {title} {Understanding quantum
  phase transitions}}}\ (\bibinfo  {publisher} {CRC press},\ \bibinfo {year}
  {2010})\BibitemShut {NoStop}%
\bibitem [{\citenamefont {Ren}\ \emph {et~al.}(2020)\citenamefont {Ren},
  \citenamefont {You},\ and\ \citenamefont {Ole\ifmmode~\acute{s}\else
  \'{s}\fi{}}}]{PhysRevB.102.024425}%
  \BibitemOpen
  \bibfield  {author} {\bibinfo {author} {\bibfnamefont {J.}~\bibnamefont
  {Ren}}, \bibinfo {author} {\bibfnamefont {W.-L.}\ \bibnamefont {You}}, \ and\
  \bibinfo {author} {\bibfnamefont {A.~M.}\ \bibnamefont
  {Ole\ifmmode~\acute{s}\else \'{s}\fi{}}},\ }\href {\doibase
  10.1103/PhysRevB.102.024425} {\bibfield  {journal} {\bibinfo  {journal}
  {Phys. Rev. B}\ }\textbf {\bibinfo {volume} {102}},\ \bibinfo {pages}
  {024425} (\bibinfo {year} {2020})}\BibitemShut {NoStop}%
\bibitem [{\citenamefont {Majumdar}\ and\ \citenamefont
  {Ghosh}(1969)}]{ckm1969}%
  \BibitemOpen
  \bibfield  {author} {\bibinfo {author} {\bibfnamefont {C.~K.}\ \bibnamefont
  {Majumdar}}\ and\ \bibinfo {author} {\bibfnamefont {D.~K.}\ \bibnamefont
  {Ghosh}},\ }\href {\doibase 10.1063/1.1664978} {\bibfield  {journal}
  {\bibinfo  {journal} {J. Math. Phys.}\ }\textbf {\bibinfo {volume} {10}},\
  \bibinfo {pages} {1388} (\bibinfo {year} {1969})},\ \Eprint
  {http://arxiv.org/abs/https://doi.org/10.1063/1.1664978}
  {https://doi.org/10.1063/1.1664978} \BibitemShut {NoStop}%
\bibitem [{\citenamefont {Thouless}(2014)}]{thouless2014quantum}%
  \BibitemOpen
  \bibfield  {author} {\bibinfo {author} {\bibfnamefont {D.~J.}\ \bibnamefont
  {Thouless}},\ }\href@noop {} {\emph {\bibinfo {title} {The quantum mechanics
  of many-body systems}}}\ (\bibinfo  {publisher} {Courier Corporation},\
  \bibinfo {year} {2014})\BibitemShut {NoStop}%
\bibitem [{\citenamefont {Dutton}\ \emph {et~al.}(2012)\citenamefont {Dutton},
  \citenamefont {Kumar}, \citenamefont {Mourigal}, \citenamefont {Soos},
  \citenamefont {Wen}, \citenamefont {Broholm}, \citenamefont {Andersen},
  \citenamefont {Huang}, \citenamefont {Zbiri}, \citenamefont {Toft-Petersen},\
  and\ \citenamefont {Cava}}]{dutton2012108}%
  \BibitemOpen
  \bibfield  {author} {\bibinfo {author} {\bibfnamefont {S.~E.}\ \bibnamefont
  {Dutton}}, \bibinfo {author} {\bibfnamefont {M.}~\bibnamefont {Kumar}},
  \bibinfo {author} {\bibfnamefont {M.}~\bibnamefont {Mourigal}}, \bibinfo
  {author} {\bibfnamefont {Z.~G.}\ \bibnamefont {Soos}}, \bibinfo {author}
  {\bibfnamefont {J.-J.}\ \bibnamefont {Wen}}, \bibinfo {author} {\bibfnamefont
  {C.~L.}\ \bibnamefont {Broholm}}, \bibinfo {author} {\bibfnamefont {N.~H.}\
  \bibnamefont {Andersen}}, \bibinfo {author} {\bibfnamefont {Q.}~\bibnamefont
  {Huang}}, \bibinfo {author} {\bibfnamefont {M.}~\bibnamefont {Zbiri}},
  \bibinfo {author} {\bibfnamefont {R.}~\bibnamefont {Toft-Petersen}}, \ and\
  \bibinfo {author} {\bibfnamefont {R.~J.}\ \bibnamefont {Cava}},\ }\href
  {\doibase 10.1103/PhysRevLett.108.187206} {\bibfield  {journal} {\bibinfo
  {journal} {Phys. Rev. Lett.}\ }\textbf {\bibinfo {volume} {108}},\ \bibinfo
  {pages} {187206} (\bibinfo {year} {2012})}\BibitemShut {NoStop}%
\bibitem [{\citenamefont {Zhitomirsky}\ and\ \citenamefont
  {Tsunetsugu}(2010)}]{zhitomirsky2010}%
  \BibitemOpen
  \bibfield  {author} {\bibinfo {author} {\bibfnamefont {M.~E.}\ \bibnamefont
  {Zhitomirsky}}\ and\ \bibinfo {author} {\bibfnamefont {H.}~\bibnamefont
  {Tsunetsugu}},\ }\href {\doibase 10.1209/0295-5075/92/37001} {\bibfield
  {journal} {\bibinfo  {journal} {E. Phys. Lett.}\ }\textbf {\bibinfo {volume}
  {92}},\ \bibinfo {pages} {37001} (\bibinfo {year} {2010})}\BibitemShut
  {NoStop}%
\bibitem [{\citenamefont {Chubukov}(1991)}]{chubukov1991}%
  \BibitemOpen
  \bibfield  {author} {\bibinfo {author} {\bibfnamefont {A.~V.}\ \bibnamefont
  {Chubukov}},\ }\href {\doibase 10.1103/PhysRevB.44.4693} {\bibfield
  {journal} {\bibinfo  {journal} {Phys. Rev. B}\ }\textbf {\bibinfo {volume}
  {44}},\ \bibinfo {pages} {4693} (\bibinfo {year} {1991})}\BibitemShut
  {NoStop}%
\bibitem [{\citenamefont {Furukawa}\ \emph {et~al.}(2012)\citenamefont
  {Furukawa}, \citenamefont {Sato}, \citenamefont {Onoda},\ and\ \citenamefont
  {Furusaki}}]{furukawa2012}%
  \BibitemOpen
  \bibfield  {author} {\bibinfo {author} {\bibfnamefont {S.}~\bibnamefont
  {Furukawa}}, \bibinfo {author} {\bibfnamefont {M.}~\bibnamefont {Sato}},
  \bibinfo {author} {\bibfnamefont {S.}~\bibnamefont {Onoda}}, \ and\ \bibinfo
  {author} {\bibfnamefont {A.}~\bibnamefont {Furusaki}},\ }\href {\doibase
  10.1103/PhysRevB.86.094417} {\bibfield  {journal} {\bibinfo  {journal} {Phys.
  Rev. B}\ }\textbf {\bibinfo {volume} {86}},\ \bibinfo {pages} {094417}
  (\bibinfo {year} {2012})}\BibitemShut {NoStop}%
\bibitem [{\citenamefont {Parvej}\ and\ \citenamefont
  {Kumar}(2017)}]{parvej2017}%
  \BibitemOpen
  \bibfield  {author} {\bibinfo {author} {\bibfnamefont {A.}~\bibnamefont
  {Parvej}}\ and\ \bibinfo {author} {\bibfnamefont {M.}~\bibnamefont {Kumar}},\
  }\href {\doibase 10.1103/PhysRevB.96.054413} {\bibfield  {journal} {\bibinfo
  {journal} {Phys. Rev. B}\ }\textbf {\bibinfo {volume} {96}},\ \bibinfo
  {pages} {054413} (\bibinfo {year} {2017})}\BibitemShut {NoStop}%
\bibitem [{\citenamefont {Saha}\ \emph {et~al.}(2020)\citenamefont {Saha},
  \citenamefont {Roy}, \citenamefont {Kumar},\ and\ \citenamefont
  {Soos}}]{saha2020modeling}%
  \BibitemOpen
  \bibfield  {author} {\bibinfo {author} {\bibfnamefont {S.~K.}\ \bibnamefont
  {Saha}}, \bibinfo {author} {\bibfnamefont {M.~S.}\ \bibnamefont {Roy}},
  \bibinfo {author} {\bibfnamefont {M.}~\bibnamefont {Kumar}}, \ and\ \bibinfo
  {author} {\bibfnamefont {Z.~G.}\ \bibnamefont {Soos}},\ }\href {\doibase
  10.1103/PhysRevB.101.054411} {\bibfield  {journal} {\bibinfo  {journal}
  {Phys. Rev. B}\ }\textbf {\bibinfo {volume} {101}},\ \bibinfo {pages}
  {054411} (\bibinfo {year} {2020})}\BibitemShut {NoStop}%
\bibitem [{\citenamefont {Maiti}\ and\ \citenamefont
  {Kumar}(2019)}]{maiti2019quantum}%
  \BibitemOpen
  \bibfield  {author} {\bibinfo {author} {\bibfnamefont {D.}~\bibnamefont
  {Maiti}}\ and\ \bibinfo {author} {\bibfnamefont {M.}~\bibnamefont {Kumar}},\
  }\href {\doibase 10.1103/PhysRevB.100.245118} {\bibfield  {journal} {\bibinfo
   {journal} {Phys. Rev. B}\ }\textbf {\bibinfo {volume} {100}},\ \bibinfo
  {pages} {245118} (\bibinfo {year} {2019})}\BibitemShut {NoStop}%
\bibitem [{\citenamefont {Bethe}(1931)}]{bethe1931theorie}%
  \BibitemOpen
  \bibfield  {author} {\bibinfo {author} {\bibfnamefont {H.}~\bibnamefont
  {Bethe}},\ }\href {https://doi.org/10.1007/BF01341708} {\bibfield  {journal}
  {\bibinfo  {journal} {Z. f{\"u}r Physik}\ }\textbf {\bibinfo {volume} {71}},\
  \bibinfo {pages} {205} (\bibinfo {year} {1931})}\BibitemShut {NoStop}%
\bibitem [{\citenamefont {Haldane}(1980)}]{haldane1980general}%
  \BibitemOpen
  \bibfield  {author} {\bibinfo {author} {\bibfnamefont {F.~D.~M.}\
  \bibnamefont {Haldane}},\ }\href {\doibase 10.1103/PhysRevLett.45.1358}
  {\bibfield  {journal} {\bibinfo  {journal} {Phys. Rev. Lett.}\ }\textbf
  {\bibinfo {volume} {45}},\ \bibinfo {pages} {1358} (\bibinfo {year}
  {1980})}\BibitemShut {NoStop}%
\bibitem [{\citenamefont {Takahashi}(1969)}]{takahashi1969magnetization}%
  \BibitemOpen
  \bibfield  {author} {\bibinfo {author} {\bibfnamefont {M.}~\bibnamefont
  {Takahashi}},\ }\href {\doibase 10.1143/PTP.42.1098} {\bibfield  {journal}
  {\bibinfo  {journal} {Prog. Theoretical Physics}\ }\textbf {\bibinfo {volume}
  {42}},\ \bibinfo {pages} {1098} (\bibinfo {year} {1969})}\BibitemShut
  {NoStop}%
\bibitem [{\citenamefont {Shiba}(1972)}]{shiba1972magnetic}%
  \BibitemOpen
  \bibfield  {author} {\bibinfo {author} {\bibfnamefont {H.}~\bibnamefont
  {Shiba}},\ }\href {\doibase 10.1103/PhysRevB.6.930} {\bibfield  {journal}
  {\bibinfo  {journal} {Phys. Rev. B}\ }\textbf {\bibinfo {volume} {6}},\
  \bibinfo {pages} {930} (\bibinfo {year} {1972})}\BibitemShut {NoStop}%
\bibitem [{\citenamefont {Affleck}\ and\ \citenamefont
  {Marston}(1988)}]{affleck1988}%
  \BibitemOpen
  \bibfield  {author} {\bibinfo {author} {\bibfnamefont {I.}~\bibnamefont
  {Affleck}}\ and\ \bibinfo {author} {\bibfnamefont {J.~B.}\ \bibnamefont
  {Marston}},\ }\href {\doibase 10.1103/PhysRevB.37.3774} {\bibfield  {journal}
  {\bibinfo  {journal} {Phys. Rev. B}\ }\textbf {\bibinfo {volume} {37}},\
  \bibinfo {pages} {3774} (\bibinfo {year} {1988})}\BibitemShut {NoStop}%
\bibitem [{\citenamefont {Marston}\ and\ \citenamefont
  {Affleck}(1989)}]{marston1989}%
  \BibitemOpen
  \bibfield  {author} {\bibinfo {author} {\bibfnamefont {J.~B.}\ \bibnamefont
  {Marston}}\ and\ \bibinfo {author} {\bibfnamefont {I.}~\bibnamefont
  {Affleck}},\ }\href {\doibase 10.1103/PhysRevB.39.11538} {\bibfield
  {journal} {\bibinfo  {journal} {Phys. Rev. B}\ }\textbf {\bibinfo {volume}
  {39}},\ \bibinfo {pages} {11538} (\bibinfo {year} {1989})}\BibitemShut
  {NoStop}%
\bibitem [{\citenamefont {Giamarchi}(2003)}]{giamarchi2003quantum}%
  \BibitemOpen
  \bibfield  {author} {\bibinfo {author} {\bibfnamefont {T.}~\bibnamefont
  {Giamarchi}},\ }\href@noop {} {\emph {\bibinfo {title} {Quantum physics in
  one dimension}}},\ Vol.\ \bibinfo {volume} {121}\ (\bibinfo  {publisher}
  {Clarendon press},\ \bibinfo {year} {2003})\BibitemShut {NoStop}%
\bibitem [{\citenamefont {Shankar}(1994)}]{shankar1994renormalization}%
  \BibitemOpen
  \bibfield  {author} {\bibinfo {author} {\bibfnamefont {R.}~\bibnamefont
  {Shankar}},\ }\href {\doibase 10.1103/RevModPhys.66.129} {\bibfield
  {journal} {\bibinfo  {journal} {Rev. Mod. Phys.}\ }\textbf {\bibinfo {volume}
  {66}},\ \bibinfo {pages} {129} (\bibinfo {year} {1994})}\BibitemShut
  {NoStop}%
\bibitem [{\citenamefont {Soos}\ and\ \citenamefont
  {Ramasesha}(1984)}]{soos1984vb}%
  \BibitemOpen
  \bibfield  {author} {\bibinfo {author} {\bibfnamefont {Z.~G.}\ \bibnamefont
  {Soos}}\ and\ \bibinfo {author} {\bibfnamefont {S.}~\bibnamefont
  {Ramasesha}},\ }\href {\doibase 10.1103/PhysRevB.29.5410} {\bibfield
  {journal} {\bibinfo  {journal} {Phys. Rev. B}\ }\textbf {\bibinfo {volume}
  {29}},\ \bibinfo {pages} {5410} (\bibinfo {year} {1984})}\BibitemShut
  {NoStop}%
\bibitem [{\citenamefont {Soos}\ and\ \citenamefont
  {Ramasesha}(1983)}]{soos1983spin}%
  \BibitemOpen
  \bibfield  {author} {\bibinfo {author} {\bibfnamefont {Z.~G.}\ \bibnamefont
  {Soos}}\ and\ \bibinfo {author} {\bibfnamefont {S.}~\bibnamefont
  {Ramasesha}},\ }\href {\doibase 10.1103/PhysRevLett.51.2374} {\bibfield
  {journal} {\bibinfo  {journal} {Phys. Rev. Lett.}\ }\textbf {\bibinfo
  {volume} {51}},\ \bibinfo {pages} {2374} (\bibinfo {year}
  {1983})}\BibitemShut {NoStop}%
\bibitem [{\citenamefont {White}(1992)}]{white-prl92}%
  \BibitemOpen
  \bibfield  {author} {\bibinfo {author} {\bibfnamefont {S.~R.}\ \bibnamefont
  {White}},\ }\href {\doibase 10.1103/PhysRevLett.69.2863} {\bibfield
  {journal} {\bibinfo  {journal} {Phys. Rev. Lett.}\ }\textbf {\bibinfo
  {volume} {69}},\ \bibinfo {pages} {2863} (\bibinfo {year}
  {1992})}\BibitemShut {NoStop}%
\bibitem [{\citenamefont {White}(1993)}]{white-prb93}%
  \BibitemOpen
  \bibfield  {author} {\bibinfo {author} {\bibfnamefont {S.~R.}\ \bibnamefont
  {White}},\ }\href {\doibase 10.1103/PhysRevB.48.10345} {\bibfield  {journal}
  {\bibinfo  {journal} {Phys. Rev. B}\ }\textbf {\bibinfo {volume} {48}},\
  \bibinfo {pages} {10345} (\bibinfo {year} {1993})}\BibitemShut {NoStop}%
\bibitem [{\citenamefont {Schollw\"ock}(2005)}]{schollwock_dmrg}%
  \BibitemOpen
  \bibfield  {author} {\bibinfo {author} {\bibfnamefont {U.}~\bibnamefont
  {Schollw\"ock}},\ }\href {\doibase 10.1103/RevModPhys.77.259} {\bibfield
  {journal} {\bibinfo  {journal} {Rev. Mod. Phys.}\ }\textbf {\bibinfo {volume}
  {77}},\ \bibinfo {pages} {259} (\bibinfo {year} {2005})}\BibitemShut
  {NoStop}%
\bibitem [{\citenamefont {Hallberg}(2006)}]{karen2006}%
  \BibitemOpen
  \bibfield  {author} {\bibinfo {author} {\bibfnamefont {K.~A.}\ \bibnamefont
  {Hallberg}},\ }\href {https://doi.org/10.1080/00018730600766432} {\bibfield
  {journal} {\bibinfo  {journal} {Adv. Phys.}\ }\textbf {\bibinfo {volume}
  {55}},\ \bibinfo {pages} {477} (\bibinfo {year} {2006})}\BibitemShut
  {NoStop}%
\bibitem [{\citenamefont {Sandvik}(2010)}]{sandvik2010}%
  \BibitemOpen
  \bibfield  {author} {\bibinfo {author} {\bibfnamefont {A.~W.}\ \bibnamefont
  {Sandvik}},\ }\href {\doibase 10.1063/1.3518900} {\bibfield  {journal}
  {\bibinfo  {journal} {AIP Conf. Proc.}\ }\textbf {\bibinfo {volume} {1297}},\
  \bibinfo {pages} {135} (\bibinfo {year} {2010})}\BibitemShut {NoStop}%
\bibitem [{\citenamefont {Sandvik}\ and\ \citenamefont
  {Kurkij\"arvi}(1991)}]{qmcex1}%
  \BibitemOpen
  \bibfield  {author} {\bibinfo {author} {\bibfnamefont {A.~W.}\ \bibnamefont
  {Sandvik}}\ and\ \bibinfo {author} {\bibfnamefont {J.}~\bibnamefont
  {Kurkij\"arvi}},\ }\href {\doibase 10.1103/PhysRevB.43.5950} {\bibfield
  {journal} {\bibinfo  {journal} {Phys. Rev. B}\ }\textbf {\bibinfo {volume}
  {43}},\ \bibinfo {pages} {5950} (\bibinfo {year} {1991})}\BibitemShut
  {NoStop}%
\bibitem [{\citenamefont {Henelius}\ and\ \citenamefont
  {Sandvik}(2000)}]{henelius2000}%
  \BibitemOpen
  \bibfield  {author} {\bibinfo {author} {\bibfnamefont {P.}~\bibnamefont
  {Henelius}}\ and\ \bibinfo {author} {\bibfnamefont {A.~W.}\ \bibnamefont
  {Sandvik}},\ }\href {\doibase 10.1103/PhysRevB.62.1102} {\bibfield  {journal}
  {\bibinfo  {journal} {Phys. Rev. B}\ }\textbf {\bibinfo {volume} {62}},\
  \bibinfo {pages} {1102} (\bibinfo {year} {2000})}\BibitemShut {NoStop}%
\bibitem [{\citenamefont {Carrasquilla}\ and\ \citenamefont
  {Melko}(2017)}]{Carrasquilla}%
  \BibitemOpen
  \bibfield  {author} {\bibinfo {author} {\bibfnamefont {J.}~\bibnamefont
  {Carrasquilla}}\ and\ \bibinfo {author} {\bibfnamefont {R.}~\bibnamefont
  {Melko}},\ }\href {\doibase 10.1038/nphys4035} {\bibfield  {journal}
  {\bibinfo  {journal} {Nature Physics}\ }\textbf {\bibinfo {volume} {13}}
  (\bibinfo {year} {2017}),\ 10.1038/nphys4035}\BibitemShut {NoStop}%
\bibitem [{\citenamefont {Torlai}\ and\ \citenamefont
  {Melko}(2016)}]{torlai2016learning}%
  \BibitemOpen
  \bibfield  {author} {\bibinfo {author} {\bibfnamefont {G.}~\bibnamefont
  {Torlai}}\ and\ \bibinfo {author} {\bibfnamefont {R.~G.}\ \bibnamefont
  {Melko}},\ }\href {\doibase 10.1103/PhysRevB.94.165134} {\bibfield  {journal}
  {\bibinfo  {journal} {Phys. Rev. B}\ }\textbf {\bibinfo {volume} {94}},\
  \bibinfo {pages} {165134} (\bibinfo {year} {2016})}\BibitemShut {NoStop}%
\bibitem [{\citenamefont {Beach}\ \emph {et~al.}(2018)\citenamefont {Beach},
  \citenamefont {Golubeva},\ and\ \citenamefont {Melko}}]{beach2018machine}%
  \BibitemOpen
  \bibfield  {author} {\bibinfo {author} {\bibfnamefont {M.~J.~S.}\
  \bibnamefont {Beach}}, \bibinfo {author} {\bibfnamefont {A.}~\bibnamefont
  {Golubeva}}, \ and\ \bibinfo {author} {\bibfnamefont {R.~G.}\ \bibnamefont
  {Melko}},\ }\href {\doibase 10.1103/PhysRevB.97.045207} {\bibfield  {journal}
  {\bibinfo  {journal} {Phys. Rev. B}\ }\textbf {\bibinfo {volume} {97}},\
  \bibinfo {pages} {045207} (\bibinfo {year} {2018})}\BibitemShut {NoStop}%
\bibitem [{\citenamefont {Broecker}\ \emph {et~al.}(2017)\citenamefont
  {Broecker}, \citenamefont {Carrasquilla}, \citenamefont {Melko},\ and\
  \citenamefont {Trebst}}]{broecker2017machine}%
  \BibitemOpen
  \bibfield  {author} {\bibinfo {author} {\bibfnamefont {P.}~\bibnamefont
  {Broecker}}, \bibinfo {author} {\bibfnamefont {J.}~\bibnamefont
  {Carrasquilla}}, \bibinfo {author} {\bibfnamefont {R.~G.}\ \bibnamefont
  {Melko}}, \ and\ \bibinfo {author} {\bibfnamefont {S.}~\bibnamefont
  {Trebst}},\ }\href {https://doi.org/10.1038/s41598-017-09098-0} {\bibfield
  {journal} {\bibinfo  {journal} {Scientific reports}\ }\textbf {\bibinfo
  {volume} {7}},\ \bibinfo {pages} {1} (\bibinfo {year} {2017})}\BibitemShut
  {NoStop}%
\bibitem [{\citenamefont {Shlens}(2014)}]{shlens2014tutorial}%
  \BibitemOpen
  \bibfield  {author} {\bibinfo {author} {\bibfnamefont {J.}~\bibnamefont
  {Shlens}},\ }\href@noop {} {\enquote {\bibinfo {title} {A tutorial on
  principal component analysis},}\ } (\bibinfo {year} {2014}),\ \Eprint
  {http://arxiv.org/abs/arXiv.1404.1100} {arXiv:arXiv.1404.1100 [cs.LG]}
  \BibitemShut {NoStop}%
\bibitem [{\citenamefont {Rodriguez-Nieva}\ and\ \citenamefont
  {Scheurer}(2019)}]{rodriguez2019identifying}%
  \BibitemOpen
  \bibfield  {author} {\bibinfo {author} {\bibfnamefont {J.~F.}\ \bibnamefont
  {Rodriguez-Nieva}}\ and\ \bibinfo {author} {\bibfnamefont {M.~S.}\
  \bibnamefont {Scheurer}},\ }\href {https://doi.org/10.1038/s41567-019-0512-x}
  {\bibfield  {journal} {\bibinfo  {journal} {Nature Physics}\ }\textbf
  {\bibinfo {volume} {15}},\ \bibinfo {pages} {790} (\bibinfo {year}
  {2019})}\BibitemShut {NoStop}%
\bibitem [{\citenamefont {Rem}\ \emph {et~al.}(2019)\citenamefont {Rem},
  \citenamefont {K{\"a}ming}, \citenamefont {Tarnowski}, \citenamefont
  {Asteria}, \citenamefont {Fl{\"a}schner}, \citenamefont {Becker},
  \citenamefont {Sengstock},\ and\ \citenamefont
  {Weitenberg}}]{rem2019identifying}%
  \BibitemOpen
  \bibfield  {author} {\bibinfo {author} {\bibfnamefont {B.~S.}\ \bibnamefont
  {Rem}}, \bibinfo {author} {\bibfnamefont {N.}~\bibnamefont {K{\"a}ming}},
  \bibinfo {author} {\bibfnamefont {M.}~\bibnamefont {Tarnowski}}, \bibinfo
  {author} {\bibfnamefont {L.}~\bibnamefont {Asteria}}, \bibinfo {author}
  {\bibfnamefont {N.}~\bibnamefont {Fl{\"a}schner}}, \bibinfo {author}
  {\bibfnamefont {C.}~\bibnamefont {Becker}}, \bibinfo {author} {\bibfnamefont
  {K.}~\bibnamefont {Sengstock}}, \ and\ \bibinfo {author} {\bibfnamefont
  {C.}~\bibnamefont {Weitenberg}},\ }\href
  {https://doi.org/10.1038/s41567-019-0554-0} {\bibfield  {journal} {\bibinfo
  {journal} {Nature Physics}\ }\textbf {\bibinfo {volume} {15}},\ \bibinfo
  {pages} {917} (\bibinfo {year} {2019})}\BibitemShut {NoStop}%
\bibitem [{\citenamefont {Costa}\ \emph {et~al.}(2017)\citenamefont {Costa},
  \citenamefont {Hu}, \citenamefont {Bai}, \citenamefont {Scalettar},\ and\
  \citenamefont {Singh}}]{PhysRevB.96.195138}%
  \BibitemOpen
  \bibfield  {author} {\bibinfo {author} {\bibfnamefont {N.~C.}\ \bibnamefont
  {Costa}}, \bibinfo {author} {\bibfnamefont {W.}~\bibnamefont {Hu}}, \bibinfo
  {author} {\bibfnamefont {Z.~J.}\ \bibnamefont {Bai}}, \bibinfo {author}
  {\bibfnamefont {R.~T.}\ \bibnamefont {Scalettar}}, \ and\ \bibinfo {author}
  {\bibfnamefont {R.~R.~P.}\ \bibnamefont {Singh}},\ }\href {\doibase
  10.1103/PhysRevB.96.195138} {\bibfield  {journal} {\bibinfo  {journal} {Phys.
  Rev. B}\ }\textbf {\bibinfo {volume} {96}},\ \bibinfo {pages} {195138}
  (\bibinfo {year} {2017})}\BibitemShut {NoStop}%
\bibitem [{\citenamefont {Rosten}\ and\ \citenamefont
  {Drummond}(2006)}]{rosten2006machine}%
  \BibitemOpen
  \bibfield  {author} {\bibinfo {author} {\bibfnamefont {E.}~\bibnamefont
  {Rosten}}\ and\ \bibinfo {author} {\bibfnamefont {T.}~\bibnamefont
  {Drummond}},\ }in\ \href {https://doi.org/10.1007/11744023_34} {\emph
  {\bibinfo {booktitle} {Computer Vision -- ECCV 2006}}},\ \bibinfo {editor}
  {edited by\ \bibinfo {editor} {\bibfnamefont {A.}~\bibnamefont {Leonardis}},
  \bibinfo {editor} {\bibfnamefont {H.}~\bibnamefont {Bischof}}, \ and\
  \bibinfo {editor} {\bibfnamefont {A.}~\bibnamefont {Pinz}}}\ (\bibinfo
  {publisher} {Springer Berlin Heidelberg},\ \bibinfo {address} {Berlin,
  Heidelberg},\ \bibinfo {year} {2006})\ pp.\ \bibinfo {pages}
  {430--443}\BibitemShut {NoStop}%
\bibitem [{\citenamefont {Zoph}\ \emph {et~al.}(2018)\citenamefont {Zoph},
  \citenamefont {Vasudevan}, \citenamefont {Shlens},\ and\ \citenamefont
  {Le}}]{zoph2018learning}%
  \BibitemOpen
  \bibfield  {author} {\bibinfo {author} {\bibfnamefont {B.}~\bibnamefont
  {Zoph}}, \bibinfo {author} {\bibfnamefont {V.}~\bibnamefont {Vasudevan}},
  \bibinfo {author} {\bibfnamefont {J.}~\bibnamefont {Shlens}}, \ and\ \bibinfo
  {author} {\bibfnamefont {Q.~V.}\ \bibnamefont {Le}},\ }\href@noop {}
  {\enquote {\bibinfo {title} {Learning transferable architectures for scalable
  image recognition},}\ } (\bibinfo {year} {2018})\BibitemShut {NoStop}%
\bibitem [{\citenamefont {Bishop}(2006)}]{bishop2006pattern}%
  \BibitemOpen
  \bibfield  {author} {\bibinfo {author} {\bibfnamefont {C.~M.}\ \bibnamefont
  {Bishop}},\ }\href@noop {} {\enquote {\bibinfo {title} {Pattern
  recognition},}\ } (\bibinfo {year} {2006})\BibitemShut {NoStop}%
\bibitem [{\citenamefont {Liben-Nowell}\ and\ \citenamefont
  {Kleinberg}(2007)}]{liben2007link}%
  \BibitemOpen
  \bibfield  {author} {\bibinfo {author} {\bibfnamefont {D.}~\bibnamefont
  {Liben-Nowell}}\ and\ \bibinfo {author} {\bibfnamefont {J.}~\bibnamefont
  {Kleinberg}},\ }\href {https://doi.org/10.1002/asi.20591} {\bibfield
  {journal} {\bibinfo  {journal} {J. American. Society. Info. Sci. Tech.}\
  }\textbf {\bibinfo {volume} {58}},\ \bibinfo {pages} {1019–1031} (\bibinfo
  {year} {2007})}\BibitemShut {NoStop}%
\bibitem [{\citenamefont {Islam}\ \emph {et~al.}(2018)\citenamefont {Islam},
  \citenamefont {Kabir}, \citenamefont {Ahmed}, \citenamefont {Kamal},
  \citenamefont {Wang},\ and\ \citenamefont {Ulhaq}}]{islam2018depression}%
  \BibitemOpen
  \bibfield  {author} {\bibinfo {author} {\bibfnamefont {M.~R.}\ \bibnamefont
  {Islam}}, \bibinfo {author} {\bibfnamefont {M.~A.}\ \bibnamefont {Kabir}},
  \bibinfo {author} {\bibfnamefont {A.}~\bibnamefont {Ahmed}}, \bibinfo
  {author} {\bibfnamefont {A.~R.~M.}\ \bibnamefont {Kamal}}, \bibinfo {author}
  {\bibfnamefont {H.}~\bibnamefont {Wang}}, \ and\ \bibinfo {author}
  {\bibfnamefont {A.}~\bibnamefont {Ulhaq}},\ }\href
  {https://pubmed.ncbi.nlm.nih.gov/30186594} {\bibfield  {journal} {\bibinfo
  {journal} {Health info. science and systems}\ }\textbf {\bibinfo {volume}
  {6}},\ \bibinfo {pages} {1} (\bibinfo {year} {2018})}\BibitemShut {NoStop}%
\bibitem [{\citenamefont {Galán-García}\ \emph {et~al.}(2015)\citenamefont
  {Galán-García}, \citenamefont {Puerta}, \citenamefont {Gómez},
  \citenamefont {Santos},\ and\ \citenamefont {Bringas}}]{galan2016supervised}%
  \BibitemOpen
  \bibfield  {author} {\bibinfo {author} {\bibfnamefont {P.}~\bibnamefont
  {Galán-García}}, \bibinfo {author} {\bibfnamefont {J.~G. d.~l.}\
  \bibnamefont {Puerta}}, \bibinfo {author} {\bibfnamefont {C.~L.}\
  \bibnamefont {Gómez}}, \bibinfo {author} {\bibfnamefont {I.}~\bibnamefont
  {Santos}}, \ and\ \bibinfo {author} {\bibfnamefont {P.~G.}\ \bibnamefont
  {Bringas}},\ }\href {\doibase 10.1093/jigpal/jzv048} {\bibfield  {journal}
  {\bibinfo  {journal} {Logic J. IGPL}\ }\textbf {\bibinfo {volume} {24}},\
  \bibinfo {pages} {42} (\bibinfo {year} {2015})}\BibitemShut {NoStop}%
\bibitem [{\citenamefont {Choi}\ and\ \citenamefont
  {Lim}(2020)}]{choimachine2020}%
  \BibitemOpen
  \bibfield  {author} {\bibinfo {author} {\bibfnamefont {J.-A.}\ \bibnamefont
  {Choi}}\ and\ \bibinfo {author} {\bibfnamefont {K.}~\bibnamefont {Lim}},\
  }\href {\doibase https://doi.org/10.1016/j.icte.2020.04.012} {\bibfield
  {journal} {\bibinfo  {journal} {ICT Express}\ }\textbf {\bibinfo {volume}
  {6}},\ \bibinfo {pages} {175} (\bibinfo {year} {2020})}\BibitemShut {NoStop}%
\bibitem [{\citenamefont {Bose}\ and\ \citenamefont
  {Mahapatra}(2001)}]{bose2001business}%
  \BibitemOpen
  \bibfield  {author} {\bibinfo {author} {\bibfnamefont {I.}~\bibnamefont
  {Bose}}\ and\ \bibinfo {author} {\bibfnamefont {R.~K.}\ \bibnamefont
  {Mahapatra}},\ }\href {\doibase
  https://doi.org/10.1016/S0378-7206(01)00091-X} {\bibfield  {journal}
  {\bibinfo  {journal} {Information \& Management}\ }\textbf {\bibinfo {volume}
  {39}},\ \bibinfo {pages} {211} (\bibinfo {year} {2001})}\BibitemShut
  {NoStop}%
\bibitem [{\citenamefont {Rajkomar}\ \emph {et~al.}(2019)\citenamefont
  {Rajkomar}, \citenamefont {Dean},\ and\ \citenamefont
  {Kohane}}]{rajkomar2019machine}%
  \BibitemOpen
  \bibfield  {author} {\bibinfo {author} {\bibfnamefont {A.}~\bibnamefont
  {Rajkomar}}, \bibinfo {author} {\bibfnamefont {J.}~\bibnamefont {Dean}}, \
  and\ \bibinfo {author} {\bibfnamefont {I.}~\bibnamefont {Kohane}},\ }\href
  {https://pubmed.ncbi.nlm.nih.gov/30943338/} {\bibfield  {journal} {\bibinfo
  {journal} {New England J. Medicine}\ }\textbf {\bibinfo {volume} {380}},\
  \bibinfo {pages} {1347} (\bibinfo {year} {2019})}\BibitemShut {NoStop}%
\bibitem [{\citenamefont {Eskov}\ \emph {et~al.}(2019)\citenamefont {Eskov},
  \citenamefont {Pyatin}, \citenamefont {Eskov},\ and\ \citenamefont
  {Ilyashenko}}]{eskov2019heuristic}%
  \BibitemOpen
  \bibfield  {author} {\bibinfo {author} {\bibfnamefont {V.~M.}\ \bibnamefont
  {Eskov}}, \bibinfo {author} {\bibfnamefont {V.~F.}\ \bibnamefont {Pyatin}},
  \bibinfo {author} {\bibfnamefont {V.}~\bibnamefont {Eskov}}, \ and\ \bibinfo
  {author} {\bibfnamefont {L.}~\bibnamefont {Ilyashenko}},\ }\href {\doibase
  10.1134/S0006350919020064} {\bibfield  {journal} {\bibinfo  {journal}
  {Biophysics}\ }\textbf {\bibinfo {volume} {64}},\ \bibinfo {pages} {293}
  (\bibinfo {year} {2019})}\BibitemShut {NoStop}%
\bibitem [{\citenamefont {Ding}\ \emph {et~al.}(2018)\citenamefont {Ding},
  \citenamefont {Chen}, \citenamefont {Cooper}, \citenamefont {Young},\ and\
  \citenamefont {Lu}}]{ding2018precision}%
  \BibitemOpen
  \bibfield  {author} {\bibinfo {author} {\bibfnamefont {M.~Q.}\ \bibnamefont
  {Ding}}, \bibinfo {author} {\bibfnamefont {L.}~\bibnamefont {Chen}}, \bibinfo
  {author} {\bibfnamefont {G.~F.}\ \bibnamefont {Cooper}}, \bibinfo {author}
  {\bibfnamefont {J.~D.}\ \bibnamefont {Young}}, \ and\ \bibinfo {author}
  {\bibfnamefont {X.}~\bibnamefont {Lu}},\ }\href {\doibase 10.1158/1541-7786}
  {\bibfield  {journal} {\bibinfo  {journal} {Molecular Cancer Research}\
  }\textbf {\bibinfo {volume} {16}},\ \bibinfo {pages} {269} (\bibinfo {year}
  {2018})}\BibitemShut {NoStop}%
\bibitem [{\citenamefont {Albertsson}\ \emph {et~al.}(2018)\citenamefont
  {Albertsson}, \citenamefont {Altoe}, \citenamefont {Anderson}, \citenamefont
  {Andrews},\ and\ \citenamefont {et. al.}}]{albertsson2018machine}%
  \BibitemOpen
  \bibfield  {author} {\bibinfo {author} {\bibfnamefont {K.}~\bibnamefont
  {Albertsson}}, \bibinfo {author} {\bibfnamefont {P.}~\bibnamefont {Altoe}},
  \bibinfo {author} {\bibfnamefont {D.}~\bibnamefont {Anderson}}, \bibinfo
  {author} {\bibfnamefont {M.}~\bibnamefont {Andrews}}, \ and\ \bibinfo
  {author} {\bibnamefont {et. al.}},\ }\href {\doibase
  10.1088/1742-6596/1085/2/022008} {\bibfield  {journal} {\bibinfo  {journal}
  {J. Phys.: Conf. Series}\ }\textbf {\bibinfo {volume} {1085}},\ \bibinfo
  {pages} {022008} (\bibinfo {year} {2018})}\BibitemShut {NoStop}%
\bibitem [{\citenamefont {Baldi}\ \emph {et~al.}(2014)\citenamefont {Baldi},
  \citenamefont {Sadowski},\ and\ \citenamefont
  {Whiteson}}]{baldi2014searching}%
  \BibitemOpen
  \bibfield  {author} {\bibinfo {author} {\bibfnamefont {P.}~\bibnamefont
  {Baldi}}, \bibinfo {author} {\bibfnamefont {P.}~\bibnamefont {Sadowski}}, \
  and\ \bibinfo {author} {\bibfnamefont {D.}~\bibnamefont {Whiteson}},\ }\href
  {\doibase 10.1038/ncomms5308} {\bibfield  {journal} {\bibinfo  {journal}
  {Nature communications}\ }\textbf {\bibinfo {volume} {5}},\ \bibinfo {pages}
  {1} (\bibinfo {year} {2014})}\BibitemShut {NoStop}%
\bibitem [{\citenamefont {Guo}\ \emph {et~al.}(2016)\citenamefont {Guo},
  \citenamefont {Liu}, \citenamefont {Oerlemans}, \citenamefont {Lao},
  \citenamefont {Wu},\ and\ \citenamefont {Lew}}]{guo2016deep}%
  \BibitemOpen
  \bibfield  {author} {\bibinfo {author} {\bibfnamefont {Y.}~\bibnamefont
  {Guo}}, \bibinfo {author} {\bibfnamefont {Y.}~\bibnamefont {Liu}}, \bibinfo
  {author} {\bibfnamefont {A.}~\bibnamefont {Oerlemans}}, \bibinfo {author}
  {\bibfnamefont {S.}~\bibnamefont {Lao}}, \bibinfo {author} {\bibfnamefont
  {S.}~\bibnamefont {Wu}}, \ and\ \bibinfo {author} {\bibfnamefont {M.~S.}\
  \bibnamefont {Lew}},\ }\href {\doibase
  https://doi.org/10.1016/j.neucom.2015.09.116} {\bibfield  {journal} {\bibinfo
   {journal} {Neurocomputing}\ }\textbf {\bibinfo {volume} {187}},\ \bibinfo
  {pages} {27} (\bibinfo {year} {2016})}\BibitemShut {NoStop}%
\bibitem [{\citenamefont {LeCun}\ \emph {et~al.}(2015)\citenamefont {LeCun},
  \citenamefont {Bengio},\ and\ \citenamefont {Hinton}}]{lecun2015deep}%
  \BibitemOpen
  \bibfield  {author} {\bibinfo {author} {\bibfnamefont {Y.}~\bibnamefont
  {LeCun}}, \bibinfo {author} {\bibfnamefont {Y.}~\bibnamefont {Bengio}}, \
  and\ \bibinfo {author} {\bibfnamefont {G.}~\bibnamefont {Hinton}},\ }\href
  {\doibase 10.1038/nature14539} {\bibfield  {journal} {\bibinfo  {journal}
  {Nature}\ }\textbf {\bibinfo {volume} {521}},\ \bibinfo {pages} {436}
  (\bibinfo {year} {2015})}\BibitemShut {NoStop}%
\bibitem [{\citenamefont {Greplova}\ \emph {et~al.}(2020)\citenamefont
  {Greplova}, \citenamefont {Valenti}, \citenamefont {Boschung}, \citenamefont
  {Schäfer}, \citenamefont {Lörch},\ and\ \citenamefont
  {Huber}}]{greplova2020unsupervised}%
  \BibitemOpen
  \bibfield  {author} {\bibinfo {author} {\bibfnamefont {E.}~\bibnamefont
  {Greplova}}, \bibinfo {author} {\bibfnamefont {A.}~\bibnamefont {Valenti}},
  \bibinfo {author} {\bibfnamefont {G.}~\bibnamefont {Boschung}}, \bibinfo
  {author} {\bibfnamefont {F.}~\bibnamefont {Schäfer}}, \bibinfo {author}
  {\bibfnamefont {N.}~\bibnamefont {Lörch}}, \ and\ \bibinfo {author}
  {\bibfnamefont {S.~D.}\ \bibnamefont {Huber}},\ }\href {\doibase
  10.1088/1367-2630/ab7771} {\bibfield  {journal} {\bibinfo  {journal} {New
  Journal of Physics}\ }\textbf {\bibinfo {volume} {22}},\ \bibinfo {pages}
  {045003} (\bibinfo {year} {2020})}\BibitemShut {NoStop}%
\bibitem [{\citenamefont {Jolliffe}(2002)}]{jolliffe2002pca}%
  \BibitemOpen
  \bibfield  {author} {\bibinfo {author} {\bibfnamefont {I.~T.}\ \bibnamefont
  {Jolliffe}},\ }\href@noop {} {\bibfield  {journal} {\bibinfo  {journal}
  {Principal component analysis}\ ,\ \bibinfo {pages} {167}} (\bibinfo {year}
  {2002})}\BibitemShut {NoStop}%
\bibitem [{\citenamefont {Jolliffe}\ and\ \citenamefont
  {Cadima}(2016)}]{jolliffe2016pca}%
  \BibitemOpen
  \bibfield  {author} {\bibinfo {author} {\bibfnamefont {I.~T.}\ \bibnamefont
  {Jolliffe}}\ and\ \bibinfo {author} {\bibfnamefont {J.}~\bibnamefont
  {Cadima}},\ }\href {\doibase 10.1098/rsta.2015.0202} {\bibfield  {journal}
  {\bibinfo  {journal} {Phil. Trans. Royal Society A: Math, Phys. and Eng.
  Sciences}\ }\textbf {\bibinfo {volume} {374}},\ \bibinfo {pages} {20150202}
  (\bibinfo {year} {2016})}\BibitemShut {NoStop}%
\bibitem [{\citenamefont {Woloshyn}(2019)}]{woloshyn2019learning}%
  \BibitemOpen
  \bibfield  {author} {\bibinfo {author} {\bibfnamefont {R.~M.}\ \bibnamefont
  {Woloshyn}},\ }\href@noop {} {\  (\bibinfo {year} {2019})},\ \Eprint
  {http://arxiv.org/abs/1905.08220} {arXiv:1905.08220 [cond-mat.stat-mech]}
  \BibitemShut {NoStop}%
\bibitem [{\citenamefont {Hu}\ \emph {et~al.}(2017)\citenamefont {Hu},
  \citenamefont {Singh},\ and\ \citenamefont {Scalettar}}]{PhysRevE.95.062122}%
  \BibitemOpen
  \bibfield  {author} {\bibinfo {author} {\bibfnamefont {W.}~\bibnamefont
  {Hu}}, \bibinfo {author} {\bibfnamefont {R.~R.~P.}\ \bibnamefont {Singh}}, \
  and\ \bibinfo {author} {\bibfnamefont {R.~T.}\ \bibnamefont {Scalettar}},\
  }\href {\doibase 10.1103/PhysRevE.95.062122} {\bibfield  {journal} {\bibinfo
  {journal} {Phys. Rev. E}\ }\textbf {\bibinfo {volume} {95}},\ \bibinfo
  {pages} {062122} (\bibinfo {year} {2017})}\BibitemShut {NoStop}%
\bibitem [{\citenamefont {Wang}(2016)}]{PhysRevB.94.195105}%
  \BibitemOpen
  \bibfield  {author} {\bibinfo {author} {\bibfnamefont {L.}~\bibnamefont
  {Wang}},\ }\href {\doibase 10.1103/PhysRevB.94.195105} {\bibfield  {journal}
  {\bibinfo  {journal} {Phys. Rev. B}\ }\textbf {\bibinfo {volume} {94}},\
  \bibinfo {pages} {195105} (\bibinfo {year} {2016})}\BibitemShut {NoStop}%
\bibitem [{\citenamefont {Wang}\ and\ \citenamefont
  {Zhai}(2017)}]{PhysRevB.96.144432}%
  \BibitemOpen
  \bibfield  {author} {\bibinfo {author} {\bibfnamefont {C.}~\bibnamefont
  {Wang}}\ and\ \bibinfo {author} {\bibfnamefont {H.}~\bibnamefont {Zhai}},\
  }\href {\doibase 10.1103/PhysRevB.96.144432} {\bibfield  {journal} {\bibinfo
  {journal} {Phys. Rev. B}\ }\textbf {\bibinfo {volume} {96}},\ \bibinfo
  {pages} {144432} (\bibinfo {year} {2017})}\BibitemShut {NoStop}%
\bibitem [{\citenamefont {Soos}\ \emph {et~al.}(2016)\citenamefont {Soos},
  \citenamefont {Parvej},\ and\ \citenamefont {Kumar}}]{soos2016}%
  \BibitemOpen
  \bibfield  {author} {\bibinfo {author} {\bibfnamefont {Z.~G.}\ \bibnamefont
  {Soos}}, \bibinfo {author} {\bibfnamefont {A.}~\bibnamefont {Parvej}}, \ and\
  \bibinfo {author} {\bibfnamefont {M.}~\bibnamefont {Kumar}},\ }\href
  {\doibase 10.1088/0953-8984/28/17/175603} {\bibfield  {journal} {\bibinfo
  {journal} {J. Phys.: Condens. Matter}\ }\textbf {\bibinfo {volume} {28}},\
  \bibinfo {pages} {175603} (\bibinfo {year} {2016})}\BibitemShut {NoStop}%
\bibitem [{\citenamefont {Okamoto}\ and\ \citenamefont
  {Nomura}(1992)}]{okamoto1992}%
  \BibitemOpen
  \bibfield  {author} {\bibinfo {author} {\bibfnamefont {K.}~\bibnamefont
  {Okamoto}}\ and\ \bibinfo {author} {\bibfnamefont {K.}~\bibnamefont
  {Nomura}},\ }\href {\doibase https://doi.org/10.1016/0375-9601(92)90823-5}
  {\bibfield  {journal} {\bibinfo  {journal} {Phys. Lett. A}\ }\textbf
  {\bibinfo {volume} {169}},\ \bibinfo {pages} {433} (\bibinfo {year}
  {1992})}\BibitemShut {NoStop}%
\bibitem [{\citenamefont {Sirker}\ \emph {et~al.}(2011)\citenamefont {Sirker},
  \citenamefont {Krivnov}, \citenamefont {Dmitriev}, \citenamefont {Herzog},
  \citenamefont {Janson}, \citenamefont {Nishimoto}, \citenamefont
  {Drechsler},\ and\ \citenamefont {Richter}}]{sirker2011}%
  \BibitemOpen
  \bibfield  {author} {\bibinfo {author} {\bibfnamefont {J.}~\bibnamefont
  {Sirker}}, \bibinfo {author} {\bibfnamefont {V.~Y.}\ \bibnamefont {Krivnov}},
  \bibinfo {author} {\bibfnamefont {D.~V.}\ \bibnamefont {Dmitriev}}, \bibinfo
  {author} {\bibfnamefont {A.}~\bibnamefont {Herzog}}, \bibinfo {author}
  {\bibfnamefont {O.}~\bibnamefont {Janson}}, \bibinfo {author} {\bibfnamefont
  {S.}~\bibnamefont {Nishimoto}}, \bibinfo {author} {\bibfnamefont {S.-L.}\
  \bibnamefont {Drechsler}}, \ and\ \bibinfo {author} {\bibfnamefont
  {J.}~\bibnamefont {Richter}},\ }\href {\doibase 10.1103/PhysRevB.84.144403}
  {\bibfield  {journal} {\bibinfo  {journal} {Phys. Rev. B}\ }\textbf {\bibinfo
  {volume} {84}},\ \bibinfo {pages} {144403} (\bibinfo {year}
  {2011})}\BibitemShut {NoStop}%
\bibitem [{\citenamefont {Kumar}\ \emph {et~al.}(2015)\citenamefont {Kumar},
  \citenamefont {Parvej},\ and\ \citenamefont {Soos}}]{mkumar2015}%
  \BibitemOpen
  \bibfield  {author} {\bibinfo {author} {\bibfnamefont {M.}~\bibnamefont
  {Kumar}}, \bibinfo {author} {\bibfnamefont {A.}~\bibnamefont {Parvej}}, \
  and\ \bibinfo {author} {\bibfnamefont {Z.~G.}\ \bibnamefont {Soos}},\ }\href
  {\doibase 10.1088/0953-8984/27/31/316001} {\bibfield  {journal} {\bibinfo
  {journal} {J. Phys.: Condens. Matter}\ }\textbf {\bibinfo {volume} {27}},\
  \bibinfo {pages} {316001} (\bibinfo {year} {2015})}\BibitemShut {NoStop}%
\bibitem [{\citenamefont {Kumar}\ \emph {et~al.}(2013)\citenamefont {Kumar},
  \citenamefont {Ramasesha},\ and\ \citenamefont {Soos}}]{mkumar2013quantum}%
  \BibitemOpen
  \bibfield  {author} {\bibinfo {author} {\bibfnamefont {M.}~\bibnamefont
  {Kumar}}, \bibinfo {author} {\bibfnamefont {S.}~\bibnamefont {Ramasesha}}, \
  and\ \bibinfo {author} {\bibfnamefont {Z.~G.}\ \bibnamefont {Soos}},\ }\href
  {\doibase http://dx.doi.org/10.5562/cca2324} {\bibfield  {journal} {\bibinfo
  {journal} {Croatica Chemica Acta}\ }\textbf {\bibinfo {volume} {86}},\
  \bibinfo {pages} {407} (\bibinfo {year} {2013})}\BibitemShut {NoStop}%
\bibitem [{\citenamefont {Chitra}\ \emph {et~al.}(1995)\citenamefont {Chitra},
  \citenamefont {Pati}, \citenamefont {Krishnamurthy}, \citenamefont {Sen},\
  and\ \citenamefont {Ramasesha}}]{chitra1995}%
  \BibitemOpen
  \bibfield  {author} {\bibinfo {author} {\bibfnamefont {R.}~\bibnamefont
  {Chitra}}, \bibinfo {author} {\bibfnamefont {S.}~\bibnamefont {Pati}},
  \bibinfo {author} {\bibfnamefont {H.~R.}\ \bibnamefont {Krishnamurthy}},
  \bibinfo {author} {\bibfnamefont {D.}~\bibnamefont {Sen}}, \ and\ \bibinfo
  {author} {\bibfnamefont {S.}~\bibnamefont {Ramasesha}},\ }\href {\doibase
  10.1103/PhysRevB.52.6581} {\bibfield  {journal} {\bibinfo  {journal} {Phys.
  Rev. B}\ }\textbf {\bibinfo {volume} {52}},\ \bibinfo {pages} {6581}
  (\bibinfo {year} {1995})}\BibitemShut {NoStop}%
\bibitem [{\citenamefont {Kumar}\ \emph {et~al.}(2010)\citenamefont {Kumar},
  \citenamefont {Soos}, \citenamefont {Sen},\ and\ \citenamefont
  {Ramasesha}}]{mkumar2010}%
  \BibitemOpen
  \bibfield  {author} {\bibinfo {author} {\bibfnamefont {M.}~\bibnamefont
  {Kumar}}, \bibinfo {author} {\bibfnamefont {Z.~G.}\ \bibnamefont {Soos}},
  \bibinfo {author} {\bibfnamefont {D.}~\bibnamefont {Sen}}, \ and\ \bibinfo
  {author} {\bibfnamefont {S.}~\bibnamefont {Ramasesha}},\ }\href {\doibase
  10.1103/PhysRevB.81.104406} {\bibfield  {journal} {\bibinfo  {journal} {Phys.
  Rev. B}\ }\textbf {\bibinfo {volume} {81}},\ \bibinfo {pages} {104406}
  (\bibinfo {year} {2010})}\BibitemShut {NoStop}%
\bibitem [{\citenamefont {White}\ and\ \citenamefont
  {Affleck}(1996)}]{srwhite1996}%
  \BibitemOpen
  \bibfield  {author} {\bibinfo {author} {\bibfnamefont {S.~R.}\ \bibnamefont
  {White}}\ and\ \bibinfo {author} {\bibfnamefont {I.}~\bibnamefont
  {Affleck}},\ }\href {\doibase 10.1103/PhysRevB.54.9862} {\bibfield  {journal}
  {\bibinfo  {journal} {Phys. Rev. B}\ }\textbf {\bibinfo {volume} {54}},\
  \bibinfo {pages} {9862} (\bibinfo {year} {1996})}\BibitemShut {NoStop}%
\end{thebibliography}%
\end{document}